\documentclass[aps,pra,twocolumn,footinbib,superscriptaddress]{revtex4-2} 
	
	\usepackage{mathtools,amsmath}
	\usepackage{booktabs}
	\usepackage{graphicx} 
	\usepackage[breaklinks=true,colorlinks,citecolor=blue,linkcolor=blue,urlcolor=blue]{hyperref}
	\usepackage[notrig]{physics}
	\usepackage{bbold}
	\usepackage[scientific-notation=true]{siunitx}
	\usepackage[dvipsnames]{xcolor}
	\usepackage{enumerate}
	\usepackage{amssymb}
	
	\usepackage[normalem]{ulem}
	
	\renewcommand{\Tr}{\mathop{\text{Tr}}\nolimits}
	\definecolor{mygray}{gray}{0.6}

\begin{document}
\title{Model-free optimization of power/efficiency tradeoffs in quantum thermal machines using reinforcement learning}

\author{Paolo A. Erdman}
\email{p.erdman@fu-berlin.de}
\affiliation{Freie Universit{\" a}t Berlin, Department of Mathematics and Computer Science, Arnimallee 6, 14195 Berlin, Germany}

\author{Frank No{\'e}}
\email{frank.noe@fu-berlin.de}
\affiliation{Microsoft Research AI4Science, Karl-Liebknecht Str. 32, 10178 Berlin, Germany}
\affiliation{Freie Universit{\" a}t Berlin, Department of Mathematics and Computer Science, Arnimallee 6, 14195 Berlin, Germany}
\affiliation{Freie Universit{\" a}t Berlin, Department of Physics, Arnimallee 6, 14195 Berlin, Germany}
\affiliation{Rice University, Department of Chemistry, Houston, TX 77005, USA}

\begin{abstract}
A quantum thermal machine is an open quantum system that enables the conversion between heat and work at the micro or nano-scale. Optimally controlling such out-of-equilibrium systems is a crucial yet challenging task with applications to quantum technologies and devices. 
We introduce a general model-free framework based on Reinforcement Learning to identify out-of-equilibrium thermodynamic cycles that are Pareto optimal trade-offs between power and efficiency for quantum heat engines and refrigerators. The method does not require any knowledge of the quantum thermal machine, nor of the system model, nor of the quantum state. Instead, it only observes the heat fluxes, so it is both applicable to simulations and experimental devices. 
We test our method on a model of an experimentally realistic refrigerator based on a superconducting qubit, and on a heat engine based on a quantum harmonic oscillator. In both cases, we identify the Pareto-front representing optimal power-efficiency tradeoffs, and the corresponding cycles. Such solutions outperform previous proposals made in the literature, such as optimized Otto cycles, reducing quantum friction.
\end{abstract}

\maketitle

\section*{Introduction}
A driving force of the research field of quantum thermodynamic is the quest of understanding and designing quantum thermal machines (QTMs), i.e. devices that convert between heat and work at the micro or nanoscale exploiting quantum effects \cite{giazotto2006,pekola2015,vinjanampathy2016,benenti2017, binder2019}.
Such devices could be operated as heat engines, which convert heat into work, or refrigerators, that extract heat from a cold bath.
Recent experiments have measured the heat flowing across these devices \cite{ronzani2018,dutta2019, senior2020, maillet2020}, and early experimental realizations of QTMs have been reported \cite{rossnagel2016, josefsson2018,klatzow2019,lindenfels2019,maslennikov2019,peterson2019, prete2019,horne2020}.

However, the optimal control of such devices, necessary to reveal their maximum performance, is an extremely challenging task that could find application in the control of quantum technologies and devices beyond QTMs. The difficulties include: (i) having to operate in finite time, the state can be driven far from equilibrium, where the thermal properties of the system are model-specific; (ii) the optimization is a search over the space of all possible time-dependent controls, which increases exponentially with the number of time points describing the cycle; (iii) in experimental devices, often subject to undesired effects such as noise and decoherence \cite{krantz2019}, we could have a limited knowledge of the actual model describing the dynamics of the QTM. 

A further difficulty (iv) arises in QTMs, since the maximization of their performance requires a multi-objective optimization. Indeed, the two main quantities that describe the performance of a heat engine (refrigerator) are the extracted power (cooling power) and the efficiency (coefficient of performance). The optimal strategy to maximize the efficiency consists of performing reversible transformations \cite{huang1987} which are, however, infinitely slow, and thus deliver vanishing power. Conversely, maximum power is typically reached at the expense of reduced efficiency. Therefore, one must seek optimal trade-offs between the two.

The theoretical optimization of QTMs is typically carried out making restrictive assumptions on the cycle. For example, optimal strategies have been derived assuming the driving speed of the control to be slow \cite{esposito2010_prl,wang2011,avron2012,ludovico2016,cavina2017_prl,abiuso2018,scandi2019,bhandari2020,alonso2022,eglinton2022} or fast \cite{abiuso2020_prl,abiuso2020_entropy,cavina2021} compared to the thermalization time. Other approaches consists of assuming \textit{a-priori} a specific shape of the cycle structure \cite{arrachea2007,esposito2010_pre,juergens2013,campisi2015,dann2020,molitor2020,shaghaghi2022,cavaliere2022}, such as the Otto cycle \cite{feldmann1996,feldmann2000,rezek2006,quan2007,abah2012,allahverdyan2013,zhang2014,campisi2016,karimi2016,kosloff2017,watanabe2017,deffner2018,gelbwaser2018,chen2019,pekola2019,das2020}.
Shortcuts to adiabaticity  \cite{berry2009,deng2013,torrontegui2013,campo2014,cakmak2018,deng2018,funo2019,villazon2019,khait2022} and variational strategies \cite{cavina2018,suri2018,menczel2019_prb} have also been employed.

In general, aside from variational approaches, there is no guarantee that these regimes and cycles are optimal. Recently, reinforcement-learning (RL) has been used to find cycles that maximize the power of QTMs without making assumptions on the cycle structure \cite{erdman2022}. However, this approach requires a model of the system and the knowledge of the quantum state of the system, which restricts its practical applicability. This calls for the development of robust and general strategies that overcome all above-mentioned difficulties (i-iv).

We propose a RL-based method with the following properties:
(i) it finds cycles yielding near Pareto-optimal trade-offs between power and efficiency, i.e. the collection of cycles such that it is not possible to further improve either power or efficiency, without decreasing the other one.
(ii) It only requires the heat currents as input, and not the quantum state of the system.
(iii) It is completely model-free.
(iv) It does not make any assumption on the cycle structure, nor on the driving speed.
The RL method is based on the Soft Actor-Critic algorithm \cite{haarnoja2018_pmlr, haarnoja2018_arxiv_sac}, introduced in the context of robotics and video-games \cite{christodoulou2019, delalleau2019}, generalized to combined discrete and continuous actions and to optimize multiple objectives.
RL has received great attention for its success at mastering tasks beyond human-level such as playing games \cite{mnih2015,silver2017,vinyals2019}, and for robotic applications \cite{haarnoja2018_arxiv_walk}. RL has been recently used for quantum control \cite{bukov2018,an2019,dalgaard2020,mackeprang2020,schafer2020,schafer2021,porotti2021,marquardt2021,brown2021,metz2023}, outperforming previous state-of-the-art methods \cite{niu2019,zhang2019}, for fault-tolerant quantum computation \cite{fosel2018,sweke2020}, and to minimize entropy production in closed quantum systems \cite{sgroi2021}.

We prove the validity of our approach finding the full Pareto-front, i.e. the collection of all Pareto-optimal cycles describing optimal power-efficency tradeoffs, in two paradigmatic systems that have been well studied in literature: 
a refrigerator based on an experimentally realistic superconducting qubit \cite{karimi2016,ronzani2018}, and a heat engine based on a quantum harmonic oscillator \cite{rezek2006}. In both cases we find elaborate cycles that outperform previous proposal mitigating quantum friction \cite{kosloff2002,rezek2006,karimi2016,friedenberger2017, brandner2017,cavina2018,pekola2019}, i.e. the detrimental effect of the generation of coherence in the instantaneous eigenbasis during the cycle. Remarkably, we can also match the performance of cycles found with the RL method of Ref.~\cite{erdman2022} that, as opposed to our model-free approach, requires monitoring the full quantum state and only optimizes the power.

\begin{figure}[!tb]
	\centering
	\includegraphics[width=0.99\columnwidth]{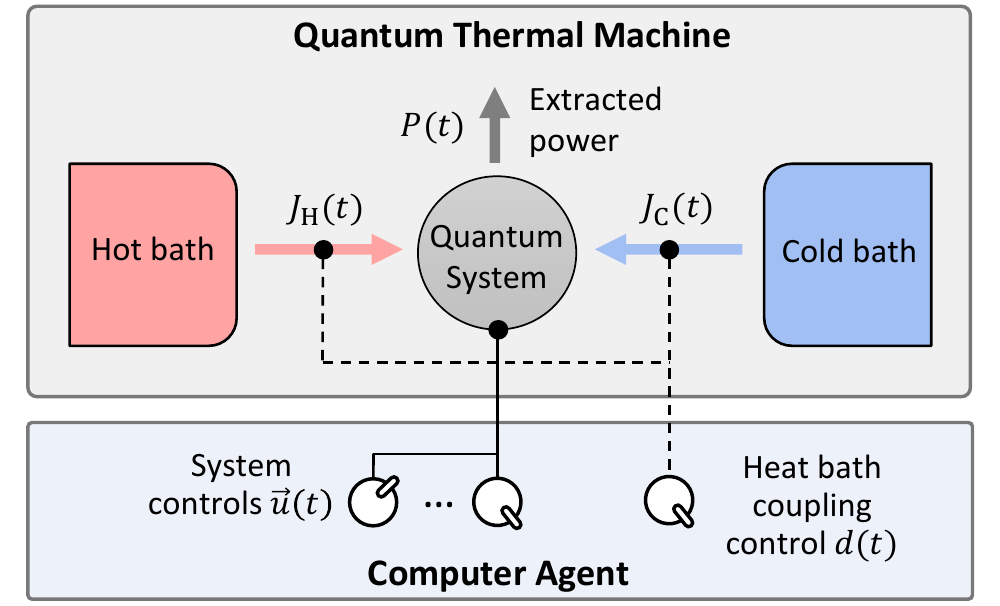}
	\caption{Schematic representation of a quantum thermal machine controlled by a computer agent. 
	A quantum system (gray circle) 
	can be coupled to a hot (cold) bath at inverse temperature $\beta_\text{H}$ ($\beta_\text{C}$), represented by the red (blue) square, 
	enabling a heat flux $J_\text{H}(t)$ ($J_\text{C}(t)$). 
	The quantum system is controlled by the computer agent through a set of experimental control parameters $\vec{u}(t)$, such as an energy gap or an oscillator frequency, that control the power exchange $P(t)$, and through a discrete control $d(t)=\{\text{Hot}, \text{Cold}, \text{None}\}$ that determines which bath is coupled to the quantum system.}
	\label{fig:setup}
\end{figure}

\subsection*{Setting: Black-box Quantum Thermal Machine}

We describe a QTM by a quantum system, acting as a ``working medium'', that can exchange heat with a hot (H) or cold (C) thermal bath characterized by inverse temperatures $\beta_\text{H}<\beta_\text{C}$ (Fig.~\ref{fig:setup}). Our method can be readily generalized to multiple baths, but we focus the description on two baths here.

We can control the evolution of the quantum system and exchange work with it through a set of time-dependent continuous control parameters $\vec{u}(t)$ that enter in the Hamiltonian $H[\vec{u}(t)]$ of the quantum system \cite{lekscha2018}, and through a discrete control $d(t)=\{\text{Hot},\text{Cold}, \text{None}\}$ that determines which bath is coupled to the system. 
$J_\text{H}(t)$ and $J_\text{C}(t)$ denote the heat flux flowing out respectively from the hot and cold bath at time $t$. 

Our method only relies on the following two assumptions:
\begin{enumerate}
    \item the RL agent can measure the heat fluxes $J_\text{C}(t)$ and $J_\text{H}(t)$ (or their averages over a time period $\Delta t$);
    \item $J_\text{C}(t)$ and $J_\text{H}(t)$ are functions of the control history ${(\vec{u}(t-T),d(t-T)),...,(\vec{u}(t),d(t))}$, where $T$ is the timescale over which the QTM remembers past controls.
\end{enumerate}
In contrast to previous work \cite{erdman2022}, the RL optimization algorithm does not require any knowledge of the microscopic model of the inner workings of the quantum system, nor of its quantum state; it is only provided with the values of the heat fluxes $J_\text{C}(t)$ and $J_\text{H}(t)$. These can be either computed from a theoretical simulation of the QTM \cite{erdman2022}, or measured directly from an experimental device whenever the energy change in the heat bath can be monitored without influencing the energetics of the quantum system (see e.g. experimental demonstrations \cite{ronzani2018,dutta2019, senior2020, maillet2020}). In this sense, our quantum system is treated as a ``black-box'', and our RL method is ``model-free''.
Any theoretical model or experimental device satisfying these requirements can be optimized by our method, including also classical stochastic thermal machines.
The timescale $T$ is finite because of energy dissipation and naturally emerges by making the minimal assumption that the coupling of the quantum system to the thermal baths drives the system towards a thermal state within some timescale $T$. 
Such a timescale can be rigorously identified e.g. within the weak system-bath coupling regime, and in the reaction coordinate framework that can describe non-Markovian and strong-coupling effects \cite{strasberg2016}. In a Markovian setting, $T$ is related to the inverse of the characteristic thermalization rate.

The thermal machines we consider are the heat engine and the refrigerator.
Up to an internal energy contribution that vanishes after each repetition of the cycle, the instantaneous power of a heat engine equals the extracted heat:
\begin{equation}
	P_{\text{heat}}(t) = J_\text{C}(t) + J_\text{H}(t),
	\label{eq_power_heat}
\end{equation}
and the cooling power of a refrigerator is:
\begin{equation}
	P_{\text{cool}}(t) = J_\text{C}(t).
	\label{eq_power_cool}
\end{equation}
The entropy production is given by
\begin{equation}
\Sigma(t) = - \beta_\text{C} J_\text{C}(t) - \beta_\text{H} J_\text{H}(t),
\end{equation}
where we neglect the contribution of the quantum system's entropy since it vanishes after each cycle.

\subsection*{Machine Learning Problem}

Our goal is to identify cycles, i.e. periodic functions $\vec{u}(t)$ and $d(t)$, that maximize a trade-off between power and efficiency \textit{on the long run}. 
Since power and efficiency cannot be simultaneously optimized, we use the concept of Pareto-optimality \cite{seoane2016,miller2019}. Pareto-optimal cycles are those where power or efficiency cannot be further increased without sacrificing the other one. The Pareto-front, defined as the collection of power-efficiency values delivered by all Pareto-optimal cycles, represents all possible optimal trade-offs. To find the Pareto-front, we define the
\textit{reward} function $r_c(t)$ as:
\begin{equation}
	r_c(t) = c \frac{P(t)}{P_0} -(1-c) \frac{\Sigma(t)}{\Sigma_0} ,
	\label{eq:r_def}
\end{equation}
where $P(t)$ is the power of a heat engine (Eq. \ref{eq_power_heat}) or cooling power of a refrigerator (Eq. \ref{eq_power_cool}), 
$P_0$, $\Sigma_0$ are reference values to normalize the power and entropy production, and
$c\in[0,1]$ is a weight that determines the trade-off between power and efficiency. 
As in Ref.~\cite{erdman2022}, we are interested in cycles that maximize the long-term performance of QTMs; we thus maximize the \textit{return} $\ev*{r_c}(t)$, where $\langle \cdot \rangle(t)$ indicates the exponential moving average of future values:
\begin{equation}
 \ev*{r_c}(t) = \kappa \int_0^\infty e^{-\kappa \tau} r_c(t+\tau)\,d\tau.
 \label{eq:continuous_return}
\end{equation}
Here $\kappa$ is the inverse of the averaging timescale, that will in practice be chosen much longer than the cycle period, 
such that $\ev*{r_c}(t)$ 
is approximately independent of $t$.

For $c=1$, we are maximizing the average power $\ev*{r_1}=\ev*{P}/P_0$. For $c=0$, we are minimizing the average entropy production $\ev*{r_0}=-\ev{\Sigma}/\Sigma_0$, which corresponds to maximizing the efficiency. For intermediate values of $c$, the maximization of $\ev*{r_c}$ describes trade-offs between power and efficiency  (see ``Optimizing the entropy production'' in Materials and Methods for details). Interestingly, if convex, it has been shown that the full Pareto-front can be identified repeating the optimization of $\ev{r_c}$ for many values of $c$ \cite{seoane2016, solon2018}.

\section*{Results}

 \begin{figure}[!tb]
	\centering
	\includegraphics[width=0.99\columnwidth]{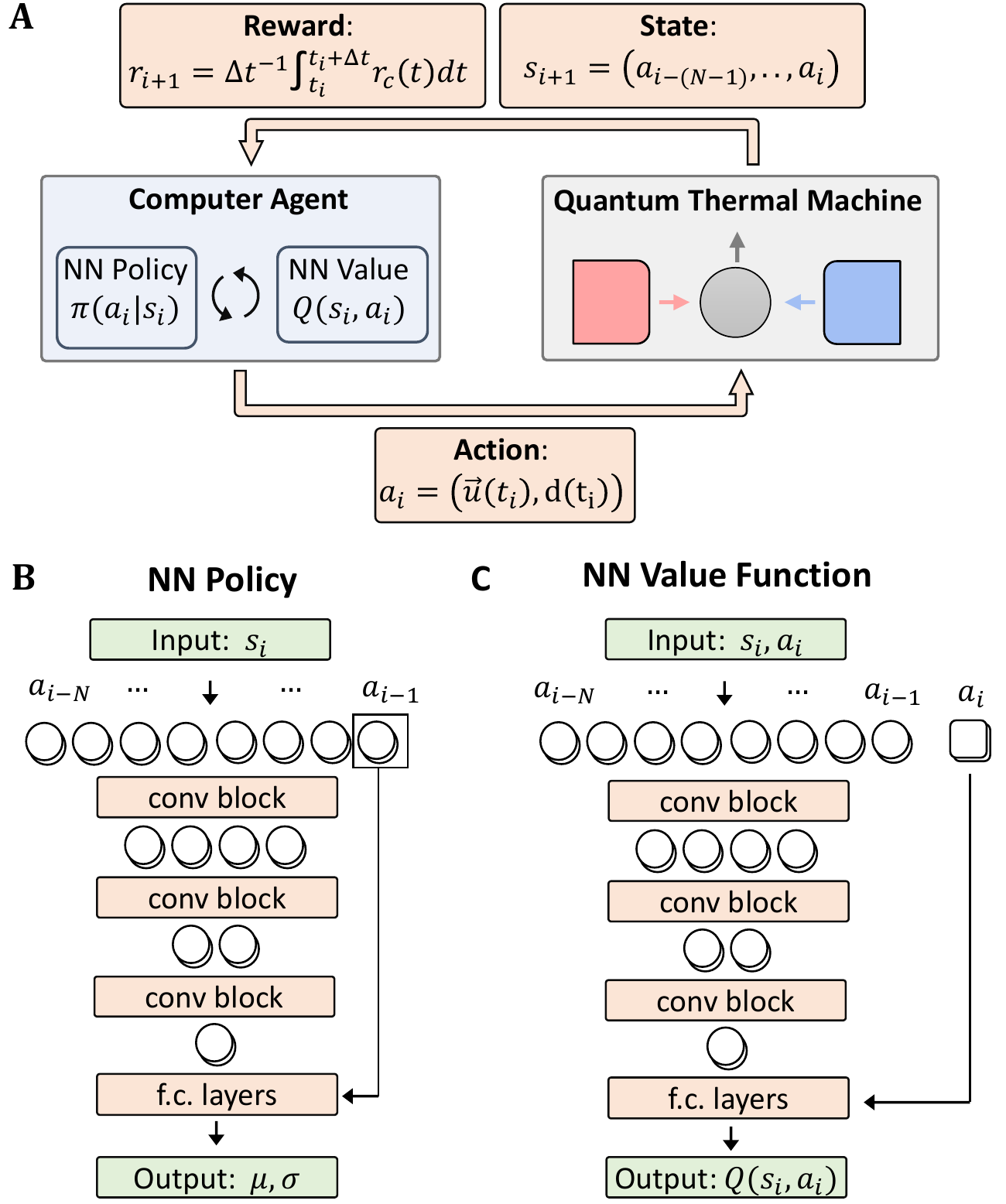}
	\caption{(A): Schematic representation of the learning process. A computer agent (blue box) chooses an action $a_i$ at time-step $i$ based on the current state $s_i$ of the QTM (gray box) through the policy function $\pi(a_i|s_i)$. The action, that encodes the control ($\vec{u}(t_i)$, $d(t_i)$), is passed to the QTM (lower  arrow). The new state $s_{i+1}$, composed of the time-series of the last $N$ actions, and the reward $r_{i+1}$ are returned to the agent (upper arrow), which uses this information to improve $\pi(a|s)$ using the soft actor-critic algorithm, which learns also the values function $Q(s,a)$. This process is reiterated until convergence of the policy. (B-C): Schematic representation of the NN architectures used to parameterize the policy (B) and the value function (C). The action time-series in $s_i$ is processed using multiple 1D convolution blocks, each one halving the length of the series. The final output is produced by fully connected (f.c.) layers.}
	\label{fig:rl}
\end{figure}

\subsection*{Deep reinforcement learning for black-box quantum thermal machines}

In RL, a \textit{computer agent} must learn to master some task by repeated interactions with some \textit{environment}.
Here we develop an RL approach where the agent maximizes the return~(\ref{eq:continuous_return}) and the environment is the QTM with its controls
(Fig.~\ref{fig:rl}A). 
To solve the RL problem computationally, we discretize time as $t_i=i \Delta t$. 
By time-discretizing the return ~(\ref{eq:continuous_return}), we obtain a discounted return whose discount factor $\gamma=\exp(-\kappa \Delta t)$ determines the averaging timescale and expresses how much we are interested in future or immediate rewards
(see ``Reinforcement Learning Implementation'' in Materials and Methods for details).

At each time step $t_i$, the agent employs a policy function $\pi(a|s)$ to choose an action $a_i = \{\vec{u}(t_i), d(t_i)\}$ based on the state $s_i$ of the environment. Here, the policy function $\pi(a|s)$ represents the probability of choosing action $a$, given that the environment is in state $s$, $\vec{u}(t)$ are the continuous controls over the quantum system, and 
$d(t_i)\in\{ \text{Hot}, \text{Cold}, \text{None} \}$ is a discrete control that selects the bath the system is coupled to. All controls are considered to be constant during time step of duration $\Delta t$. The aim of RL is to learn an optimal policy function $\pi(a|s)$ that maximizes the return.

In order to represent a black-box quantum system whose inner mechanics are unknown, we define the control history during a time interval of length $T$ as the observable state: 
\begin{equation}
    s_{i} = (a_{i-N},a_{i-N+1}, \dots, a_{i-1} ),
\end{equation}
where $N=T/\Delta t$. Therefore, the state of the quantum system is implicitly defined by the
sequence of the agent's $N$ recent actions.

To find an optimal policy we employ the soft actor-critic algorithm, that relies on learning also a value function $Q(s,a)$, generalized to a combination of discrete and continuous actions \cite{haarnoja2018_pmlr,haarnoja2018_arxiv_sac,christodoulou2019,delalleau2019}. The policy function $\pi(a|s)$ plays the role of an ``actor'' that chooses the actions to perform, while a value function $Q(s,a)$ plays the role of a ``critic'' that judges the choices made by the actor, thus providing feedback to improve the actor's behavior.
We further optimize the method for a multi-objective setting by introducing a separate critic for each objective, i.e. one value function for the power, and one for the entropy production. This allow us to vary the weight $c$ during training, thus enhancing convergence (see ``Reinforcement Learning Implementation'' in Materials and Methods for details).

We learn the functions $\pi(a|s)$ and $Q(s,a)$ using
a deep NN architecture inspired by WaveNet, an architecture that was developed for processing audio signals \cite{oord2016} (See Figs. ~\ref{fig:rl}B-C). We introduce a ``convolution block'' to efficiently process the time-series of actions defining the state $s_i$. It consists of a 1D convolution with kernel size and stride of $2$, such that it halves the length of the input. It is further equipped with a residual connection to improve trainability \cite{he2015} (see ``Reinforcement Learning Implementation'' in Materials and Methods for details). 
The policy $\pi(a_i|s_i)$ is described by a NN that takes the state $s_i$ as input, and outputs parameters $\mu$ and $\sigma$ describing the probability distribution from which action $a_i$ is sampled (Fig. ~\ref{fig:rl}B). 
The value function $Q(s_i,a_i)$ is computed by feeding $(s_i,a_i)$ into a NN, and outputting $Q(s_i,a_i)$ (Fig. ~\ref{fig:rl}C).  
Both $\pi(a_i|s_i)$ and $Q(s_i,a_i)$ process the state by feeding it through multiple convolution blocks (upper orange boxes in Figs.~\ref{fig:rl}B and \ref{fig:rl}C), each one halving the length of the time-series, such that the number of blocks and of parameters in the NN is logarithmic in $N$. Then a series of fully-connected layers produce the final output.

The policy and value functions are determined by minimizing the loss functions in Eqs.~(\ref{eq:q_loss}) and (\ref{eq:pi_loss}) using the ADAM optimization algorithm \cite{kingma2014}. The gradient of the loss functions is computed off-policy, over a batch of past experience recorded in a replay buffer, using back-propagation (see ``Reinforcement Learning Implementation'' in Materials and Methods for details).

\subsection*{Pareto-optimal cycles for a superconducting qubit refrigerator}
\begin{figure}[!tb]
	\centering
	\includegraphics[width=0.99\columnwidth]{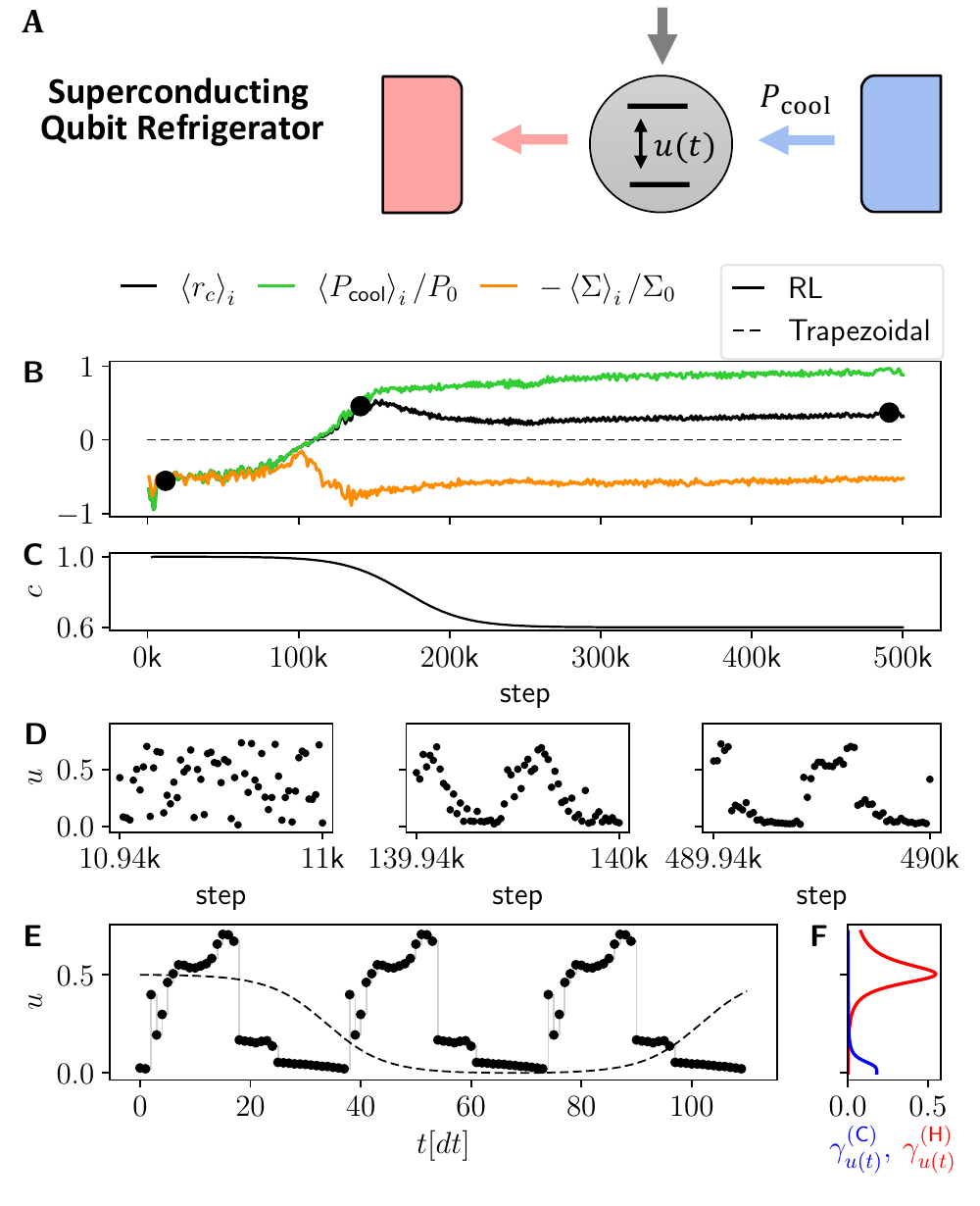}
\caption{
Training of the superconducting qubit refrigerator model to optimize $\ev*{r_c}$ at $c=0.6$. (A): Schematic representation of the energy levels of the qubit (horizontal black lines) that are controlled by $u(t)$. The gray arrow represents the input power, while the colored arrows represent the heat fluxes. (B): Return $\ev*{r_c}_i$ computed over past rewards (black curve), running average of the cooling power $\ev*{P_\text{cool}}_i/P_0$ (green curve), and of the negative entropy production $-\ev*{\Sigma}_i/\Sigma_0$ (orange curve), as a function of the training step. The dashed line represents the value of the return found optimizing the period of a smoothed trapezoidal cycle. (C): Value of the weight $c$ as a function of the step. It is varied during training from $1$ to the final value $0.6$ to improve convergence. (D): Actions chosen by the agent, represented by the value of $u$, as a function of step, zoomed around the three black circles in panel (B). (E): Final deterministic cycle found by the agent (thick black dots) and smoothed trapezoidal cycle (thin dashed line) whose return is given by the dashed line in panel (B), as a function of time. (F): coupling strength $\gamma^\text{(C)}_u$ (blue curve) and $\gamma^\text{(H)}_u$ (red curve) as a function of $u$ (on the y-axis). The parameters used for training are $N=128$, $g_\mathrm{H}=g_\mathrm{C}=1$, $\beta_\mathrm{H}=10/3$, $\beta_\mathrm{C} = 2\beta_\mathrm{H}$, $Q_\mathrm{H}=Q_\mathrm{C}=4$, $E_0=1$, $\Delta =0.12$, $\omega_\mathrm{H} = 1.028$, $\omega_\mathrm{C}=0.24$, $\mathcal{U}=[0,0.75]$, $\Delta t=0.98$, $\gamma=0.997$, $P_0 = 6.62\cdot 10^{-4}$ and $\Sigma_0=0.037$.}
	\label{fig:qubit_fridge}
\end{figure}

We first consider a refrigerator based on an experimentally realistic system: a superconducting qubit coupled to two resonant circuits that behave as heat baths \cite{karimi2016} (Fig.~\ref{fig:qubit_fridge}A). Such a system  was experimentally studied in the steady-state in Ref.~\cite{ronzani2018}. The system Hamiltonian is given by ~\cite{karimi2016,pekola2019,funo2019}:
\begin{equation}
	\hat{H}[u(t)] = - E_0\left[\Delta \hat{\sigma}_x + u(t)\hat{\sigma}_z  \right],
	\label{eq:h_fridge}
\end{equation}
where $E_0$ is a fixed energy scale, $\Delta$ characterizes the minimum gap of the system, and $u(t)$ is our control parameter. 
In this setup the coupling to the baths, described by the commonly employed Markovian master equation \cite{gorini1976,lindblad1976,breuer2002,yamaguchi2017}, 
is fixed, and cannot be controlled.
However, the qubit is resonantly coupled to the baths at different energies. 
The $u$-dependent coupling strength to the cold (hot) bath is described by the function $\gamma^{(\text{C})}_u$ ($\gamma^{(\text{H})}_u$), respectively (Fig.~\ref{fig:qubit_fridge}F). As in Ref.~\cite{funo2019}, the coupling strength is, respectively, maximal at $u=0$ ($u=1/2$), with a resonance width determined by the ``quality factor'' $Q_\text{C}$ ($Q_\text{H}$) (see ``Physical model'' in Materials and Methods for details). This allows us to choose which bath is coupled to the qubit by tuning $u(t)$.

In Fig.~\ref{fig:qubit_fridge} we show an example of our training procedure to optimize the return $\ev{r_c}$ at $c=0.6$ using $N=128$ steps determining the RL state, and varying $c$ during training from $1$ to $0.6$ (Fig.~\ref{fig:qubit_fridge}C).
In the early stages of the training, the return $\ev*{r_c}_i$, computed as in Eq.~(\ref{eq:return_def}) but over past rewards, and the running averages of the cooling power $\ev*{P_\text{cool}}_i$ and of the negative entropy production $-\ev*{\Sigma}_i$ all start off negative (Fig.~\ref{fig:qubit_fridge}B), and the corresponding actions are random (left panel of Fig.~\ref{fig:qubit_fridge}D). Indeed, initially the RL agent has no experience controlling the QTM, so random actions are performed, resulting in heating the cold bath, rather than cooling it, and in a large entropy production. However, with increasing steps, the chosen actions exhibit some structure (Fig.~\ref{fig:qubit_fridge}D), and the return $\ev*{r_c}_i$ increases (Fig.~\ref{fig:qubit_fridge}B).
While both the power and the negative entropy production initially increase together, around step 100k we see that $-\ev*{\Sigma}_i$ begins to decrease. 
This is a manifestation of the fact that power and entropy production cannot be simultaneously optimized. Indeed, the agent learns that in order to further increase the return, it must ``sacrifice'' some entropy production to produce a positive and larger cooling power. In fact, the only way to achieve positive values of $\ev*{r_c}_i$ is to have a positive cooling power, which inevitably requires producing entropy. Eventually all quantities in Fig.~\ref{fig:qubit_fridge}B reach a maximum value, and the corresponding final deterministic cycle (i.e. the cycle generated by policy switching off stochasticity, see ``Reinforcement Learning Implementation'' in Materials and Methods for details) is shown in Fig.~\ref{fig:qubit_fridge}E as thick black dots.

For the same system, Ref.~\cite{funo2019} proposed a smoothed trapezoidal cycle $u(t)$ oscillating between the resonant peaks at $u=0$ and $u=1/2$ and optimized the cycle time (Fig.~\ref{fig:qubit_fridge}E, dashed line). While this choice outperformed a sine and a trapezoidal cycle \cite{karimi2016}, the cycle found by our RL agent produces a larger return (Fig.~\ref{fig:qubit_fridge}B). 
The optimal trapezoidal cycle found for $c=0.6$ is shown in Fig.~\ref{fig:qubit_fridge}E as a dashed line
(see ``Comparing with other methods'' in Materials and Methods for details). 

\begin{figure}[!tb]
	\centering
	\includegraphics[width=0.99\columnwidth]{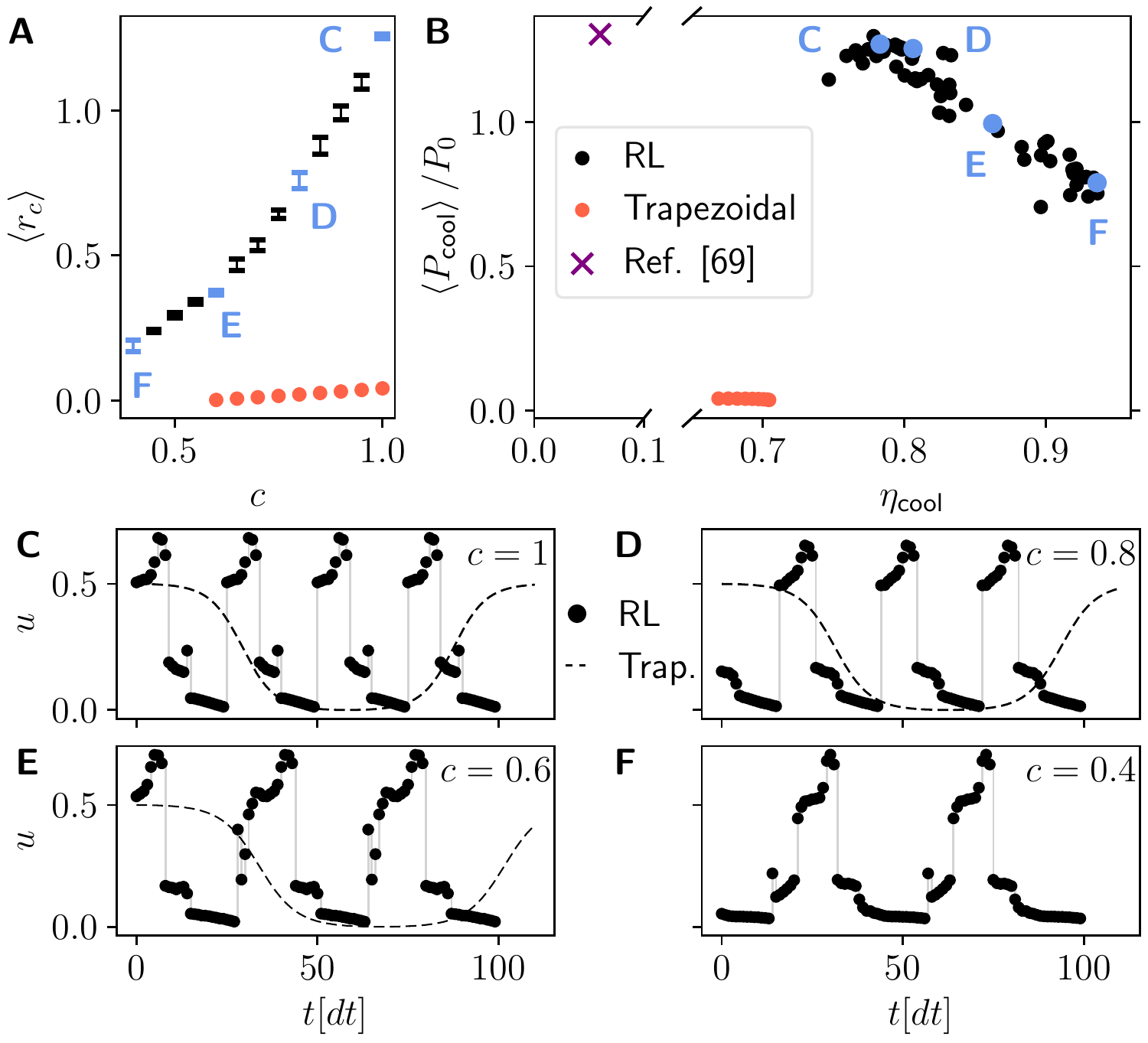}
\caption{
Results for the optimization of the superconducting qubit refrigerator model. (A): final value of the return $\ev*{r_c}$, as a function of $c$, found using the RL method (black and blue points), and optimizing the period of a trapezoidal cycle (red dots). The error bars represent the standard deviation of the return computed over $5$ independent training runs. (B): corresponding values of the final average cooling power $\ev*{P_\text{cool}}$ and of the coefficient of performance $\eta_\text{cool}$ found using the RL method (black and blue dots), optimizing the trapezoidal cycle (red dots), and using the RL method of Ref.~\cite{erdman2022} (purple cross). Results for each of the $5$ repetitions are shown as separate points to visualize the variability across multiple trainings. (C-F): final deterministic cycles identified by the RL method (thick black dots), as a function of time, corresponding to the blue points in panels (A) and (B) (respectively for $c= 1, 0.8, 0.6, 0.4$ choosing the training run with the largest return). The dashed line represents the trapezoidal cycle that maximizes the return for the same value of $c$ [not shown in panel (F) since no cycle yields a positive return]. The parameters used for training are chosen as in Fig.~\ref{fig:qubit_fridge}. }
	\label{fig:qubit_fridge_pareto}
\end{figure}

Fig.~\ref{fig:qubit_fridge_pareto} compares optimal cycles for different trade-offs between cooling power and coefficient of performance $\eta_\text{cool}$, the latter defined as the ratio between the average cooling power, and the average input power. This is achieved by repeating the optimization for various values of $c$. To demonstrate the robustness of our method, the optimization of $\ev{r_c}$ was repeated $5$ times for each choice of $c$
(variability shown with  error bars in Fig.\ref{fig:qubit_fridge_pareto}A,  and as separate points in Fig.\ref{fig:qubit_fridge_pareto}B). The RL method substantially outperforms the trapezoidal cycle by producing larger final values of the return $\ev*{r_c}$ at all values of $c$ (Fig.~\ref{fig:qubit_fridge_pareto}A), and by producing a better Pareto front (Fig.~\ref{fig:qubit_fridge_pareto}B).
The RL cycles simultaneously yield higher power by more than a factor of 10, and a larger $\eta_\text{cool}$, for any choice of the power-efficiency trade-off.
The model-free RL cycles can also deliver the same power at a substantially higher COP (roughly $10$ times larger) as compared to the cycle found with the RL method of Ref.~\cite{erdman2022}, which only optimizes the power. This is remarkable since, as opposed to the current model-free method, the method in Ref.~\cite{erdman2022} has access to the full quantum state of the system, and not only to the heat currents (see ``Comparing with other methods'' in Materials and Methods for details). This also shows that a large efficiency improvement can be achieved by sacrificing very little power. 

As expected, the period of the RL cycles increases as $c$ decreases and the priority shifts from high power to high $\eta_\text{cool}$ (Figs.~\ref{fig:qubit_fridge_pareto}C-F, black dots). However, the period is much shorter than the corresponding optimized trapezoidal cycle (dashed line), and the optimal control sequence is quite unintuitive, even going beyond the resonant point at $u=1/2$.
 As argued in \cite{karimi2016,pekola2019,funo2019}, the generation of coherence in the instantaneous eigenbasis of the quantum system, occurring because $[\hat{H}(u_1), \hat{H}(u_2)]\neq 0$ for $u_1\neq u_2$, causes power losses that increase with the speed of the cycle. We find that we can interpret the power enhancement achieved by our cycle as a mitigation of such detrimental effect: indeed, we find that trapezoidal cycles operated
at the same frequency as the RL cycle generate twice as much coherence as the RL cycles (see ``Generation of coherence'' in Materials and Methods for details). In either case, cycles with higher power tend to generate more coherence.

Given the stochastic nature of RL, we also compared the cycles obtained across the $5$ independent training runs, finding that cycles are typically quite robust, displaying only minor changes (see Fig.~\ref{fig:qubit_extra_cycles} of Methods for four cycles found in independent training runs corresponding to Figs.~\ref{fig:qubit_fridge_pareto}C-F).

\subsection*{Pareto-optimal cycles for a quantum harmonic oscillator engine}

\begin{figure}[!tb]
	\centering	\includegraphics[width=0.99\columnwidth]{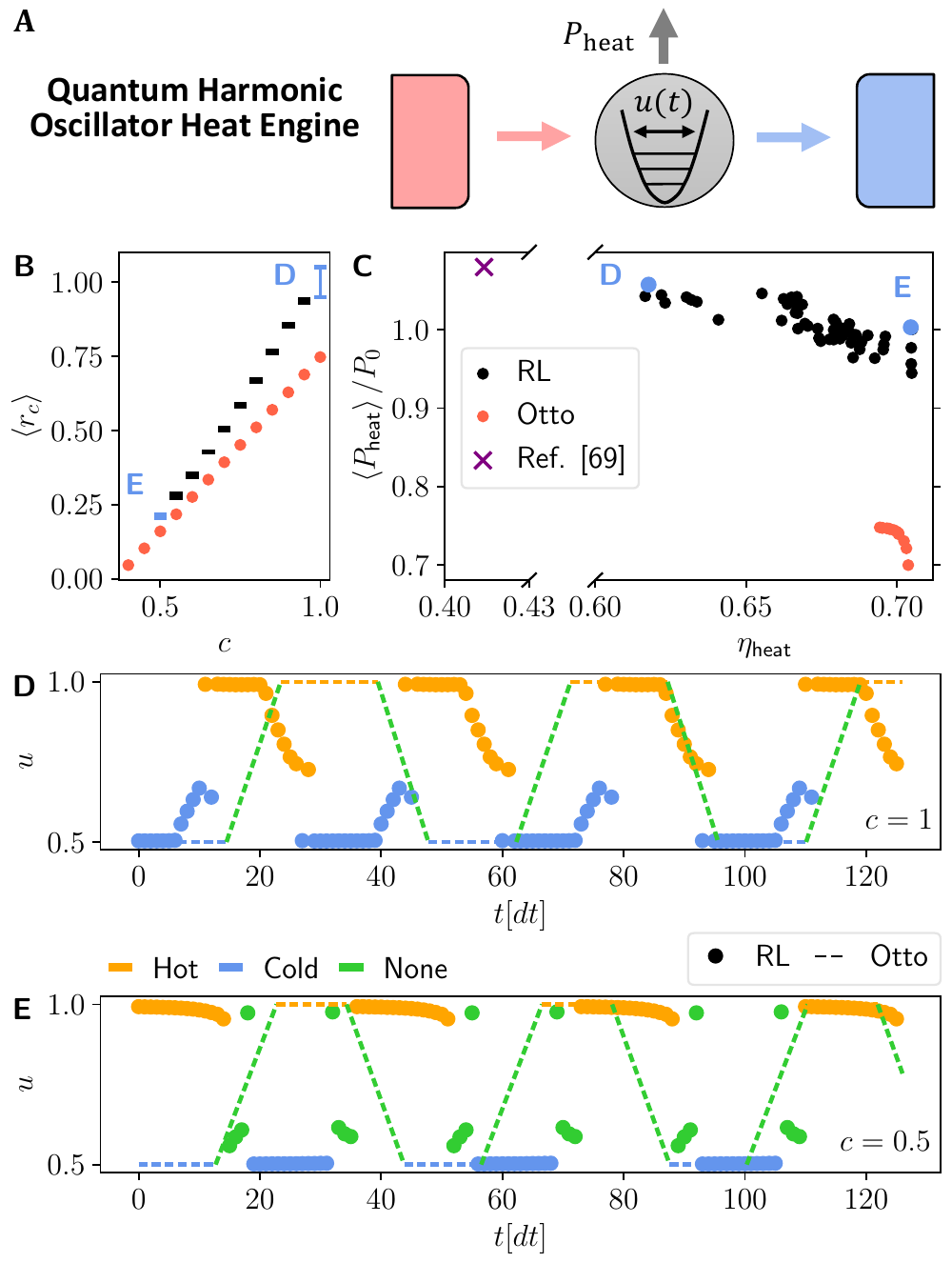}
	\caption{
	Results for the optimization of the harmonic oscillator heat engine model. (A): Schematic representation of the energy levels of the particles (black horizontal lines) trapped in a harmonic potential (parabolic curve) whose amplitude is controlled by $u(t)$. The gray arrow represents the extracted power, while the colored arrows represent the heat fluxes. (B): final value of $\ev*{r_c}$, as a function of $c$, found using the RL method (black and blue dots), and optimizing the Otto cycle (red dots). The error bars represent the standard deviation of the return computed over $5$ independent training runs. (C): corresponding values of the average power $\ev*{P_\text{heat}}/P_0$ and of the efficiency $\eta_\text{heat}$ found using the RL method (black and blue dots), optimizing the Otto cycle (red dots), and using the RL method of Ref.~\cite{erdman2022} (purple cross). Results for each of the $5$ repetitions are shown as separate points to visualize the variability across multiple trainings. (D-E): final deterministic cycle identified by the RL method (thick dots), as a function of time, corresponding to the blue points in panels (B) and (C) (respectively $c=1, 0.5$ choosing the training run with the largest return).  The color corresponds to the discrete choice $d=\{\text{Hot},\text{Cold},\text{None}\}$ (see legend). The dashed line represents the Otto cycle that maximizes the return for the same value of $c$. The parameters used for training are $N=128$, $\Gamma^{(\text{H})}=\Gamma^{(\text{C})}=0.6$, $\beta_\text{H}=0.2$, $\beta_\text{C} = 2$, $w_0=2$, $\mathcal{U}=[0.5,1]$ (to enable a fair comparison with Ref.~\cite{rezek2006}), $\Delta t=0.2$, $\gamma=0.999$, $P_0=0.175$ and $\Sigma_0=0.525$. }
	\label{fig:harmonic_engine}
\end{figure}

We now consider a heat engine based on a collection of non-interacting particles confined in a harmonic potential \cite{rezek2006} (Fig.~\ref{fig:harmonic_engine}A). The Hamiltonian is given by
\begin{equation}
	\hat{H}[u(t)] = \frac{1}{2m} \hat{p}^2 + \frac{1}{2}m (u(t)w_0)^2 \hat{q}^2,
\end{equation}
where $m$ is the mass of the system, $w_0$ is a reference frequency and $\hat{p}$ and $\hat{q}$ are the momentum and position operators. The control parameter $u(t)$ allows us to change the frequency of the oscillator. Here, at every time step we let the agent choose which bath (if any) to couple to the oscillator. The coupling to the baths, characterized by the thermalization rates $\Gamma_\alpha$, is modeled using the Lindblad master equation as in Ref.~\cite{rezek2006} (see ``Physical model'' in Materials and Methods for details). In contrast to the superconducting qubit case, $c$ is held constant during training. 

Fig.~\ref{fig:harmonic_engine} reports the results on the optimal trade-offs between extracted power and efficiency $\eta_\text{heat}$, the latter defined as the ratio between the extracted power and the input heat, in the same style of Fig.~\ref{fig:qubit_fridge_pareto}. In this setup, we compare our RL-based results to the well-known Otto cycle. The authors of Ref.~\cite{rezek2006} study this system by optimizing the switching times of an Otto cycle, i.e. the duration of each of the 4 segments, shown as a dashed lines in Figs.~\ref{fig:harmonic_engine}D-E, composing the cycle (see ``Comparing with other methods'' in Materials and Methods for details).

The RL method produces cycles with a larger return and with a better power-efficiency Pareto-front with respect to the Otto cycle (Fig.~\ref{fig:harmonic_engine}B,C). The cycles found by the RL method significantly outperforms the Otto engine in terms of delivered power.
For $c=1$, a high-power cycle is found (Fig.~\ref{fig:harmonic_engine}D and corresponding blue dots in Figs.~\ref{fig:harmonic_engine}B-C) but at the cost of a lower efficiency than the Otto cycles. 
However, at $c=0.5$, the RL method finds a cycle that matches the maximum efficiency of the Otto cycles, while delivering a $\sim 30\%$ higher power (Fig.~\ref{fig:harmonic_engine}E and corresponding blue dots in Figs.~\ref{fig:harmonic_engine}B-C)
Remarkably, our model-free RL method also finds cycles with nearly the same power as the RL method of Ref.~\cite{erdman2022}, but at almost twice the efficiency (see ``Comparing with other methods'' in Materials and Methods for details). As in Fig.~\ref{fig:qubit_fridge_pareto}, we see that a very small decrease in power can lead to a large efficiency increase.

Interestingly, as shown in Figs.~\ref{fig:harmonic_engine}D-E, the cycles found by the RL agent share many similarities with the Otto cycle: both alternate between the hot and cold bath (orange and blue portions) with a similar period. However, there are some differences: 
at $c=1$, the RL cycle ramps the value of $u$ while in contact with the bath, eliminating the unitary stroke (Fig.~\ref{fig:harmonic_engine}D). 
Instead, at $c=0.5$, the RL agent employs a unitary stroke that is quite different respect to a linear ramping of $u$ (Fig.~\ref{fig:harmonic_engine}E, green dots). As in the superconducting qubit case, the enhanced performance of the RL cycle may be interpreted as a mitigation of quantum friction \cite{kosloff2002,rezek2006}.

Also in this setup, we verified that the discovered cycles are quite robust across the $5$ independent training runs, displaying only minor changes (see Fig.~\ref{fig:harmonic_extra_cycles} of Methods for two cycles found in independent training runs corresponding to Figs.~\ref{fig:harmonic_engine}D-E).

\section*{Discussion}
We introduced a model-free framework, based on Reinforcement Learning, to discover Pareto-optimal thermodynamic cycles that describe the best possible trade-off between power and efficiency of out-of-equilibrium quantum thermal machines (heat engines and refrigerators). 
Our algorithm only requires monitoring the heat fluxes of the QTM, making it a model-free approach.
 It can therefore be used both for the theoretical optimization of known systems, and potentially for the direct optimization of experimental devices for which no model is known, and in the absence of any measurement performed on the quantum system.  Using state-of-the-art machine learning techniques, we demonstrate the validity of our method applying it to two different prototypical setups. Our black-box method discovered elaborate cycles that outperform previously proposed cycles and are on par with a previous RL method that observes the full quantum state \cite{erdman2022}. Up to minor details, the cycles found by our method are reproducible across independent training runs. Physically we find that Otto cycles, commonly studied in literature, are not generally optimal, and that optimal cycles balance a fast operation of the cycle, with the mitigation of quantum friction.

Our method paves the way for a systematic use of RL in the field of quantum thermodynamics.
Future directions include investing larger systems to uncover the impact of quantum many-body effects on the performance of QTMs, optimizing systems in the presence of noise, and optimizing trade-offs that include power fluctuations \cite{barato2015,guarnieri2019,miller2019,miller2021}.

\section*{Methods}
In this section we provide details on the optimization of the entropy production, on the reinforcement learning implementation, on the physical model used to describe the quantum thermal machines, on the training details, on the convergence of the method, on the comparison with other methods, and on the computation of the generation of coherence during the cycles. We also provide access to the full code that was used to generate the results presented in the manuscript, and the corresponding data.

~\\ \textbf{Optimizing the entropy production}

Here we discuss the relation between optimizing the power and the entropy production, or the power and the efficiency. We start by noticing that we can express the efficiency of a heat engine $\eta_\text{heat}$ and the coefficient of performance of a refrigerator $\eta_\text{cool}$ in terms of the averaged power and entropy production, i.e.
\begin{equation}
	\eta_{\nu} = \eta_{\nu}^\text{(c)}\, [1+\ev{\Sigma}/(\beta_{\nu}\ev*{P_{\nu}}) ]^{-1},
	\label{eq:eta_nu}
\end{equation}
where $\nu=\text{heat},\text{cool}$, $\eta_\text{heat}^\text{(c)}\equiv 1 -\beta_\text{H}/\beta_\text{C}$ is the Carnot efficiency, $\eta_\text{cool}^\text{(c)}\equiv \beta_\text{H}/(\beta_\text{C} - \beta_\text{H})$ is the Carnot coefficient of performance, and where we defined $\beta_\text{heat} \equiv \beta_\text{C}$ and $\beta_\text{cool} \equiv \beta_\text{C}-\beta_\text{H}$. We now show that, thanks to this dependence of $\eta_{\nu}$ on $\ev*{P_{\nu}}$ and $\ev*{\Sigma}$, if a cycle is a Pareto-optimal trade-off between high power and high efficiency, then it is also a Pareto-optimal trade-off between high power and low entropy-production up to a change of $c$. This means that if we find all optimal trade-offs between high power and low entropy-production (as we do with our method if the Pareto-front is convex), we will have necessarily also found all Pareto-optimal trade-offs between high power and high efficiency.

Mathematically, we want to prove that the cycles that maximize 
\begin{equation}
	\ev*{G_{\nu}(c)} \equiv c \ev*{P_{\nu}}  +(1-c) \eta_{\nu}
\end{equation}
for some value of $c\in[0,1]$, also maximize the return in Eq.~(\ref{eq:continuous_return}) for some (possibly different) value of $c\in[0,1]$. To simplify the proof and the notation, we consider the following two functions
\begin{equation}
\begin{aligned}
	F(a,b,\theta) &= a P(\theta) - b \Sigma(P(\theta),\eta(\theta)),\\
	G(a,b,\theta) &= a P(\theta) + b \eta(\theta),
\end{aligned}
\end{equation}
where $P(\theta)$ and $\eta(\theta)$ represent the power and efficiency of a cycle parameterized by a set of parameters $\theta$, $a> 0$ and $b> 0$ are two scalar quantities, and 
\begin{equation}
    \Sigma(P,\eta)= \frac{\eta^{\text{(c)}}_{\nu}-\eta}{\eta} \beta_{\nu}P
    \label{eq:sigma_p_eta}
\end{equation}
is obtained by inverting Eq.~(\ref{eq:eta_nu}).

We wish to prove the following. 
Given some weights $a_1>0$ and $b_1>0$, let $\theta_1$ be the value of $\theta$ that locally maximizes $G(a_1,b_1,\theta)$. Then, it is always possible to identify positive weights $a_2>0$, $b_2>0$ such that the same parameters $\theta_1$ (i.e. the same cycle) is a local maximum for $F(a_2,b_2,\theta)$. In the following, we will use that
\begin{align}
	\partial_P \Sigma &\geq 0 & \partial_\eta \Sigma &< 0,
	\label{eq:s_cond}
\end{align}
and that the Hessian $H^{(\Sigma)}$ of $\Sigma(P,\eta)$ is given by
\begin{equation}
	H^{(\Sigma)} = 
	\begin{pmatrix}
		0 & -\beta_{\nu}\frac{\eta^{\text{(c)}}_{\nu}}{\eta^2} \\
		-\beta_{\nu}\frac{\eta^{\text{(c)}}_{\nu}}{\eta^2} & 2\beta_{\nu}P\frac{\eta^{\text{(c)}}_{\nu}}{\eta^3}
	\end{pmatrix}.
	\label{eq:s_hess}
\end{equation}

Proof: by assumption, $\theta_1$ is a local maximum for $G(a_1,b_1,\theta)$. Denoting with $\partial_i$ the partial derivative in $(\theta)_i$, we thus have
\begin{equation}
	0 = \partial_i G(a_1,b_1,\theta_1) = a_1 \partial_i P(\theta_1) + b_1 \partial_i \eta(\theta_1).
	\label{eq:partial_g}
\end{equation}
Now, let us compute the derivative in $\theta$ of $F(a_2,b_2,\theta_1)$, where $a_2>0$ and $b_2>0$ are two arbitrary positive coefficients. We have
\begin{equation}
	\partial_i F(a_2,b_2,\theta_1) = ( a_2 - b_2 \partial_P \Sigma )\partial_i P(\theta_1) - (b_2\partial_\eta \Sigma)\partial_i \eta(\theta_1).
\end{equation}
Therefore, if we choose $a_2$ and $b_2$ such that 
\begin{equation}
\begin{pmatrix}
	a_1\\b_1
\end{pmatrix}
=
\begin{pmatrix}
	1 & -\partial_P \Sigma \\
	0 & -\partial_\eta \Sigma
\end{pmatrix}
\begin{pmatrix}
	a_2\\b_2
\end{pmatrix},
\label{eq:a_transf}
\end{equation}
thanks to Eq.~(\ref{eq:partial_g}) we have that 
\begin{equation}
	0 = \partial_i F(a_2,b_2,\theta_1),
\end{equation}
meaning that the same parameters $\theta_1$ that nullifies the gradient of $G$, nullifies also the gradient of $F$ at a different choice of the weights, given by Eq.~(\ref{eq:a_transf}).
The invertibility of Eq.~(\ref{eq:a_transf}) (i.e. a non-null determinant of the matrix) is guaranteed by Eq.~(\ref{eq:s_cond}). We also have to make sure that if $a_1>0$ and $b_1>0$, then also $a_2>0$ and $b_2>0$. To do this, we invert Eq.~(\ref{eq:a_transf}), finding
\begin{equation}
\begin{pmatrix}
	a_2\\b_2
\end{pmatrix}
=
\begin{pmatrix}
	1 & -\partial_P \Sigma/(\partial_\eta \Sigma) \\
	0 & -1/(\partial_\eta \Sigma)
\end{pmatrix}
\begin{pmatrix}
	a_1\\b_1
\end{pmatrix}.
\label{eq:a_inv}
\end{equation}
It is now easy to see that also the weights $a_2$ and $b_2$ are positive using Eq.~(\ref{eq:s_cond}).

To conclude the proof, we show that $\theta_1$ is a local maximum for $F(a_2,b_2,\theta)$ by showing that its Hessian is negative semi-definite. Since, by hypothesis, $\theta_1$ is a local maximum for $G(a_1,b_1,\theta)$, we have that the Hessian matrix
\begin{equation}
	H^{(G)}_{ij} \equiv \partial_{ij} G(a_1,b_1,\theta_1) = a_1 \partial_{ij}P + b_1 \partial_{ij}\eta 
\end{equation}
is negative semi-definite. We now compute the Hessian $H^{(F)}$ of $F(a_2,b_2,\theta)$ in $\theta=\theta_1$:
\begin{equation}
	H^{(F)}_{ij} = a_2 \partial_{ij}P - b_2\left[ \partial_P \Sigma\,\partial_{ij}P + \partial_\eta \Sigma\,\partial_{ij}\eta + Q_{ij}  \right],
	\label{eq:hg_def}
\end{equation}
where
\begin{equation}
	Q_{ij} =
	\begin{pmatrix}
		\partial_i P & \partial_i \eta
	\end{pmatrix}
	H^{(\Sigma)}
	\begin{pmatrix}
		\partial_j P \\ \partial_j \eta
	\end{pmatrix},
	\label{eq:qij_def}
\end{equation}
and	$H^{(\Sigma)}$ is the Hessian of $\Sigma(P,\eta)$ computed in $P(\theta_1)$ and $\eta(\theta_1)$.
Since we are interested in studying the Hessian of $F(a_2,b_2,\theta_1)$ in the special point $(a_2,b_2)$ previously identified, we substitute Eq.~(\ref{eq:a_inv}) into Eq.~(\ref{eq:hg_def}), yielding
\begin{equation}
	H^{(F)}_{ij} = H^{(G)}_{ij} + \frac{b_1}{\partial_\eta \Sigma} Q_{ij}.
\end{equation}
We now prove that $H^{(F)}_{ij}$ is negative semi-definite since it is the sum of negative semi-definite matrices. By hypothesis $H^{(G)}_{ij}$ is negative semi-definite. Recalling Eq.~(\ref{eq:s_cond}) and that $b_1 >0$, we now need to show that $Q_{ij}$ is positive semi-definite. Plugging Eq.~(\ref{eq:s_hess}) into Eq.~(\ref{eq:qij_def}) yields
\begin{equation}
	Q_{ij} = \beta_{[\nu]}\frac{\eta^{\text{(c)}}_{[\nu]}}{\eta^2}\partial_i\eta\,\partial_j\eta\,R_{ij},
\end{equation}
where
\begin{align}
	R_{ij} &\equiv 2\frac{P}{\eta} + S_{ij} + S^T_{ij}, &
	S_{ij} &= -\frac{\partial_i P}{\partial_i \eta}.
	\label{eq:r_def}
\end{align}
We now show that if $R_{ij}$ is positive semi-definite, then also  $Q_{ij}$ is positive semi-definite. By definition, $Q_{ij}$ is positive semidefinite if, for any set of coefficient $a_i$, we have that $\sum_{ij} a_i Q_{ij} a_j\geq 0$. Assuming $R_{ij}$ to be positive semi-definite, and using that $\beta_{[\nu]}, \eta^{\text{(c)}}_{[\nu]}, \eta >0$, we have 
\begin{equation}
	\sum_{ij} a_i Q_{ij} a_j = \beta_{[\nu]}\frac{\eta^{\text{(c)}}_{[\nu]}}{\eta^2} \sum_{ij} x_i R_{ij}x_j \geq 0,
\end{equation}
where we define $x_i \equiv \partial_i \eta \,a_i$. We thus have to prove the positivity of $R_{ij}$. We prove this showing that it is the sum of $3$ positive semi-definite matrices. Indeed, the first term in Eq.~(\ref{eq:r_def}), $2P/\eta$, is proportional to a matrix with $1$ in all entries. Trivially, this matrix has $1$ positive eigenvalue, and all other ones are null, so it is positive semi-definite. At last, $S_{ij}$ and its transpose have the same positivity, so we focus only on $S_{ij}$. $S_{ij}$ is a matrix with all equal columns. This means that it has all null eigenvalues, except for a single one that we denote with $\lambda$. Since the trace of a matrix is equal to the sum of the eigenvalues, we have $\lambda = \mathrm{Tr}[S]=\sum_i S_{ii}$. Using the optimality condition in Eq.~(\ref{eq:partial_g}), we see that each entry of $S$ is positive, i.e. $S_{ij}>0$. Therefore $\lambda >0$, thus $S$ is positive semi-definite, concluding the proof that $H^{(F)}_{ij}$ is negative semi-definite.

To conclude, we notice that we can always renormalize $a_2$ and $b_2$, preserving the same exact optimization problem. This way, a value of $c\in[0,1]$ can be identified.

~\\ \textbf{Reinforcement Learning Implementation}

As discussed in the main text, our goal is to maximize the return $\ev{r_c}(t)$ defined in Eq.~(\ref{eq:continuous_return}). To solve the problem within the RL framework, we discretize time as $t_i=i\Delta t$. At every time-step $t_i$, the aim of the agent is to learn an optimal policy that maximizes, in expectation, the time-discretized return $\ev*{r_c}_i$. 
The time-discrete reward and return functions are given by:
\begin{eqnarray}
    r_{i+1}&=&\Delta t^{-1} \int_{t_i}^{t_i+\Delta_t} r_c(t) \mathrm{d}t, \label{eq:ri_def}
    \\
	\ev*{r_c}_i &=& 
	(1-\gamma)\sum_{j=0}^\infty \gamma^{j} r_{i+1+j}.
	\label{eq:return_def}
\end{eqnarray}
Eq. (\ref{eq:return_def}) is the time-discrete version
of Eq. (\ref{eq:continuous_return}),
where the \textit{discount factor} $\gamma = \exp(-\kappa \Delta t)$ determines the averaging timescale and expresses how much we are interested in future or immediate rewards.

 To be precise, plugging Eq.~(\ref{eq:ri_def}) into Eq.~(\ref
  {eq:return_def}) gives $\langle r_c \rangle(t) $  (up to an irrelevant constant prefactor) only in the limit of
  $\Delta t \to 0$. However, also for finite $\Delta t$, both quantities are  time-averages of the reward, so they are equally valid definitions to describe a long-term trade-off maximization.
  
As in Ref.~\cite{erdman2022}, we use a generalization of the soft-actor critic (SAC) method, first developed for continuous actions
\cite{haarnoja2018_pmlr,haarnoja2018_arxiv_sac}, to handle a combination of discrete and continuous actions \cite{christodoulou2019,delalleau2019}. We further tune the method to stabilize the convergence in a multi-objective scenario.
We here present an overview of our implementation of SAC putting special emphasis on the differences with respect to the standard implementation. However, we refer to \cite{haarnoja2018_pmlr,haarnoja2018_arxiv_sac,christodoulou2019,delalleau2019} for additional details. Our method, implemented with PyTorch, is based on modifications and generalizations of the SAC implementation provided by Spinning Up from OpenAI \cite{spinningup2018}. All code and data to reproduce the experiments is available online (see Data Availability and Code Availability sections).

The SAC algorithm is based on policy iteration, i.e. it consists of iterating multiple times over two steps: a \textit{policy evaluation step}, and a \textit{policy improvement step}. In the policy evaluation step, the value function of the current policy is (partially) learned, whereas in the policy improvement step a better policy is learned by making use of the value function. We now describe these steps more in detail.

In typical RL problems, the optimal policy $\pi^*(s|a)$ is defined as the policy that maximizes the expected return defined in Eq.~(\ref{eq:return_def}), i.e.:
\begin{equation}
    \pi^* = \mathrm{arg}\max_\pi\, \mathop{\mathrm{E}_\pi}\limits_{ s\sim \mu_\pi}  \Big[ \sum_{k=0}^{\infty} \gamma^k \,r_{k+1} \Big| s_0 = s \Big],
    \label{eq:pi_star_rl}
\end{equation}
where $\mathrm{E}_\pi$ denotes the expectation value choosing actions according to the policy $\pi$. The initial state $s_0=s$ is sampled from $\mu_\pi$, i.e. the steady-state distribution of states that are visited by $\pi$. 
In the SAC method, balance between exploration and exploitation \cite{sutton2018} is achieved by introducing an Entropy-Regularized maximization objective. In this setting, the optimal policy $\pi^*$ is given by
\begin{equation}
    \pi^* = \mathrm{arg}\max_\pi\, \mathop{\mathrm{E}_\pi}\limits_{s\sim \mathcal{B}}\Big[ \sum_{k=0}^{\infty} \gamma^k \,\Big(r_{k+1} + \alpha H[\pi(\cdot|s_k)]  \Big) \Big| s_0 =s  \Big],
    \label{eq:pi_star}
\end{equation}
where $\alpha \geq 0$ is known as the ``temperature'' parameter that balances the trade-off between exploration and exploitation, and
\begin{equation}
    H[P] = \mathop{\mathrm{E}}\limits_{x\sim P}[ -\log P(x) ]
\end{equation}
is the entropy of the probability distribution $P$. Notice that we replaced the unknown state distribution $\mu_\pi$ with $\mathcal{B}$, which is a replay buffer populated during training by storing the observed one-step transitions $(s_k,a_k, r_{k+1}, s_{k+1})$.

Developing on Ref.~\cite{erdman2022}, we generalize such approach to a combination of discrete and continuous actions in the following way. Let us write an arbitrary action $a$ as $a = (u,d)$, where $u$ is the continuous action and $d$ is the discrete action (for simplicity, we describe the case of a single continuous action, though the generalization to multiple variables is straightforward). From now on, all functions of $a$ are also to be considered as functions of $u,d$. We decompose the joint probability distribution of the policy as
\begin{equation}
    \pi(u,d|s) = \pi_{\mathrm{D}}(d|s) \cdot \pi_{\mathrm{C}}(u|d,s),
    \label{eq:pi_decomp}
\end{equation}
where $\pi_{\mathrm{D}}(d|s)$ is the marginal probability of taking discrete action $d$, and $\pi_{\mathrm{C}}(u|d,s)$ is the conditional probability density of choosing action $u$, given action $d$ (D stands for ``discrete'', and C for ``continuous''). Notice that this decomposition is an exact identity, thus allowing us to describe correlations between the discrete and the continuous action. With this decomposition, we can write the entropy of a policy as
\begin{equation}
    H[\pi(\cdot|s)] = H^{\pi}_\text{D}(s) + H^{\pi}_\text{C}(s),
    \label{eq:h_decomposition}
\end{equation}
where 
\begin{align}
    H^{\pi}_\text{D}(s) &= H[\pi_{\mathrm{D}}(\cdot|s)], &  
    H^{\pi}_\text{C}(s) &= \sum_d \pi_{\mathrm{D}}(d|s) H[\pi_{\mathrm{C}}(\cdot|d,s)],
\end{align}
correspond respectively to the entropy contribution of the discrete (D) and continuous (C) part.
These two entropies take on values in different ranges: while the entropy of a discrete distribution with $|D|$ discrete actions is non-negative and upper bounded by $\log(|D|)$, the (differential) entropy of a continuous distribution can take on any value, including negative values (especially for peaked distributions). Therefore, we introduce a separate temperature for the discrete and continuous contributions replacing the definition of the optimal policy in Eq.~(\ref{eq:pi_star}) with
\begin{multline}
    \pi^* = \mathrm{arg}\max_\pi\, \mathop{\mathrm{E}_\pi}\limits_{s\sim \mathcal{B}}\Big[ \sum_{k=0}^{\infty} \gamma^k \,\Big(r_{k+1} + \alpha_\text{D} H^\pi_\text{D}(s_k) \\ 
     + \alpha_\text{C}H^\pi_\text{C}(s_k)  \Big) \Big| s_0 =s  \Big],
    \label{eq:pi_star_final}
\end{multline}
where $\alpha_\text{C} \geq 0$ and $\alpha_\text{D} \geq 0$ are two distinct ``temperature'' parameters. This is one of the differences with respect to Refs.~\cite{haarnoja2018_pmlr,haarnoja2018_arxiv_sac,erdman2022}.
Equation~(\ref{eq:pi_star_final}) defines our optimization objective. Accordingly, we define the value function $Q^\pi(s,a)$ of a given policy $\pi$ as
\begin{multline}
    Q^\pi(s,a) = 
    \text{E}_\pi  \Big[ r_{1} + 
    \sum_{k=1}^{\infty} \gamma^k \,\Big(r_{k+1} + \alpha_\text{D} H^\pi_\text{D}(s_k) \\
     + \alpha_\text{C}H^\pi_\text{C}(s_k)  \Big) \Big| s_0=s, a_0=a \Big].
    \label{eq:q_def}
\end{multline}
Its recursive Bellman equation therefore reads
\begin{multline}
    Q^\pi(s,a) =  \underset{s_1 \atop a_1 \sim \pi(\cdot|s_1) }{\text{E}} \Big[ r_{1} + 
    \gamma \Big( Q^\pi(s_1,a_1) + \alpha_\text{D} H^\pi_\text{D}(s_1) \\
    + \alpha_\text{C}H^\pi_\text{C}(s_1)  \Big) \Big| s_0=s, a_0=a \Big].
    \label{eq:bellman}
\end{multline}
As in Ref.~\cite{haarnoja2018_pmlr,haarnoja2018_arxiv_sac}, we parameterize $\pi_{\mathrm{C}}(u|d,s)$ as a squashed Gaussian policy, i.e. as the distribution of the variable
\begin{equation}
\begin{aligned}
 \tilde{u}(\xi|d,s) &= 
    u_\text{a} + \frac{u_\text{b} - u_\text{a}}{2}[1+ \tanh\left( \mu(d,s) + \sigma(d,s)\cdot \xi )  \right)],   
     \\  \xi &\sim \mathcal{N}(0,1),
\end{aligned}
\label{eq:u_tilda}
\end{equation}
where $\mu(d,s)$ and $\sigma(d,s)$ represent respectively the mean and standard deviation of the Gaussian distribution, $\mathcal{N}(0,1)$ is the normal distribution with zero mean and unit variance, and where we assume that $\mathcal{U}=[u_\mathrm{a},u_\mathrm{b}]$. This is the so-called reparameterization trick.

We now describe the policy evaluation step. In the SAC algorithm, we learn two value functions $Q_{\phi_i}(s,a)$ described by the learnable parameters $\phi_i$, for $i=1,2$. $Q_\phi(s,a)$ is a function approximator, e.g. a neural network. 
Since $Q_{\phi_i}(s,a)$ should satisfy the Bellman Eq.~(\ref{eq:bellman}), we define the loss function for $Q_{\phi_i}(s,a)$ as the mean square difference between the left and right hand side of Eq.~(\ref{eq:bellman}), i.e.
\begin{equation}
    L_Q(\phi_i) = \mathop{\mathrm{E}}\limits_{(s,a,r,s^\prime)\sim \mathcal{B}} \left[ ( Q_{\phi_i}(s,a) - y(r,s^\prime))^2  \right],
    \label{eq:q_loss}
\end{equation}
where 
\begin{multline}
    y(r,s^\prime) = r +  \gamma \underset{a^\prime \sim \pi(\cdot|s^\prime)}{\text{E}} \Big[ \min_{j=1,2}Q_{\phi_{\text{targ},j}}(s^\prime,a^\prime) + \alpha_\text{D} H_\text{D}(s^\prime) \\
     + \alpha_\text{C}H_\text{C}(s^\prime) \Big].
    \label{eq:y_1}
\end{multline}
Notice that in Eq.~(\ref{eq:y_1}) we replaced $Q^\pi$ with $\min_{j=1,2}Q_{\phi_{\mathrm{targ},j}}$, where $\phi_{\mathrm{targ},j}$, for $j=1,2$, are target parameters which are not updated when minimizing the loss function; instead, they are held fixed during backpropagation, and then they are updated according to Polyak averaging, i.e.
\begin{equation}
    \phi_{\mathrm{targ},i} \leftarrow \rho_\mathrm{polyak} \phi_{\mathrm{targ},i} + (1-\rho_\mathrm{polyak})\phi_{i},
\end{equation}
where $\rho_\mathrm{polyak}$ is a hyperparameter. This change was shown to improve learning \cite{haarnoja2018_pmlr,haarnoja2018_arxiv_sac}. In order to evaluate the expectation value in Eq.~(\ref{eq:y_1}), we use the decomposition in Eq.~(\ref{eq:pi_decomp}) to write
\begin{equation}
    \mathop{\mathrm{E}}\limits_{ a^\prime\sim \pi(\cdot|s^\prime)}[\cdot] =
   \sum_{{d}^\prime} \pi_{\mathrm{D}}({d}^\prime|s^\prime) \mathop{\mathrm{E}}\limits_{{u}^\prime \sim \pi_{\text{C}}(\cdot|d^\prime,s^\prime) }[\cdot] ,
   \label{eq:e_decomp}
\end{equation}
where we denote $a^\prime = (u^\prime, d^\prime)$. Plugging Eq.~(\ref{eq:e_decomp}) into Eq.~(\ref{eq:y_1}) and writing the entropies explicitly as expectation values yields
\begin{multline}
    y(r,s^\prime) = r + \gamma \sum_{{d}^\prime} \pi_{\text{D}}({d}^\prime|s^\prime) \cdot \Big(  \\
      \mathop{\mathrm{E}}\limits_{ u^\prime\sim \pi_{\text{C}}(\cdot|{d}^\prime,s^\prime)}  \left[  \min_{j=1,2}Q_{\phi_{\text{targ},j}}(s^\prime,d^\prime,u^\prime) - \alpha_\text{C} \log \pi_{\text{C}}(u^\prime|d^\prime,s^\prime) \right] \\
     - \alpha_\text{D}\log\pi_{\text{D}}(d^\prime|s^\prime) \Big).
    \label{eq:y_2}
\end{multline}

We then replace the expectation value over $u^\prime$ in Eq.~(\ref{eq:y_2}) with a single sampling $u^\prime\sim\pi_{\text{C}}(\cdot|{d}^\prime,s^\prime)$ (therefore one sampling for each discrete action) performed using Eq.~(\ref{eq:u_tilda}). This corresponds to performing a full average over the discrete action, and a single sampling of the continuous action.

We now turn to the policy improvement step. Since we introduced two separate temperatures, we cannot use the loss function introduced in Refs.~\cite{haarnoja2018_pmlr,haarnoja2018_arxiv_sac}. Therefore, we proceed in two steps. Let us define the following function
\begin{equation}
    Z_\pi(s) = - \mathop{\mathrm{E}}\limits_{{a\sim \pi(\cdot|s)}}\left[ Q^{\pi^\text{old}}(s,a)   \right] - \alpha_\text{D} H_\text{D}^\pi(s) - \alpha_\text{C} H^\pi_\text{C}(s),
    \label{eq:z_def}
\end{equation}
where $Q^{\pi^\text{old}}(s,a)$ is the value function of some given ``old policy'' $\pi^\text{old}$, and $\pi$ is an arbitrary policy. First, we prove that if a policy $\pi^\text{new}$ satisfies
\begin{equation}
    Z_{\pi^\text{new}}(s) \leq Z_{\pi^\text{old}}(s)
    \label{eq:z_prop}
\end{equation}
for all values of $s$, then $\pi^\text{new}$ is a better policy than $\pi^\text{old}$ as defined in Eq.~(\ref{eq:pi_star_final}). Next, we will use this property to define a loss function that implements the policy improvement step. 
Equation (\ref{eq:z_prop}) implies that
\begin{multline}
    \mathop{\mathrm{E}}\limits_{{a\sim \pi^\text{old}(\cdot|s)}}\left[ Q^{\pi^\text{old}}(s,a)   \right] + \alpha_\text{D} H_\text{D}^{\pi^\text{old}}(s) + \alpha_\text{C} H^{\pi^\text{old}}_\text{C}(s) \leq \\
     \mathop{\mathrm{E}}\limits_{{a\sim \pi^\text{new}(\cdot|s)}}\left[ Q^{\pi^\text{old}}(s,a)   \right] + \alpha_\text{D} H_\text{D}^{\pi^\text{new}}(s) + \alpha_\text{C} H^{\pi^\text{new}}_\text{C}(s).
     \label{eq:improvement_ineq}
\end{multline}
We now use this inequality to show that $\pi^\text{new}$ is a better policy. Starting from the Bellmann equation (\ref{eq:bellman}) for $Q^{\pi^\text{old}}$, we have Eq.~(\ref{eq:improve_proof}).
\begin{figure*}[tb!]
\begin{multline}
        Q^{\pi^\text{old}}(s,a) =  \underset{ s_1 \atop a_1 \sim \pi^\text{old}(\cdot|s_1) }{\text{E}} \Big[ r_{1} + 
    \gamma \Big( Q^{\pi^\text{old}}(s_1,a_1) + \alpha_\text{D} H^{\pi^\text{old}}_\text{D}(s_1) + \alpha_\text{C}H^{\pi^\text{old}}_\text{C}(s_1)  \Big) \Big| s_0=s, a_0=a \Big] \leq \\
    \underset{ s_1 \atop a_1 \sim \pi^\text{new}(\cdot|s_1) }{\text{E}} \Big[ r_{1} + 
    \gamma \Big( Q^{\pi^\text{old}}(s_1,a_1) + \alpha_\text{D} H^{\pi^\text{new}}_\text{D}(s_1) + \alpha_\text{C}H^{\pi^\text{new}}_\text{C}(s_1)  \Big) \Big| s_0=s, a_0=a \Big] = \\
    \underset{ s_1 \atop a_1 \sim \pi^\text{new}(\cdot|s_1) }{\text{E}} \Big[ r_{1} + 
    \gamma \Big(\alpha_\text{D} H^{\pi^\text{new}}_\text{D}(s_1) + \alpha_\text{C}H^{\pi^\text{new}}_\text{C}(s_1)  \Big) \Big| s_0=s, a_0=a \Big] +
    \gamma\underset{ s_1 \atop a_1 \sim \pi^\text{new}(\cdot|s_1) }{\text{E}} \Big[
    Q^{\pi^\text{old}}(s_1,a_1) \Big| s_0=s, a_0=a \Big] \\
    \leq \dots \leq Q^{\pi^\text{new}}(s,a).
    \label{eq:improve_proof}
\end{multline}
\end{figure*}

Using a strategy similar to that described in Refs.~\cite{sutton2018,haarnoja2018_pmlr}, in Eq.~(\ref{eq:improve_proof}) we make a repeated use of inequality (\ref{eq:improvement_ineq}) and of the Bellmann equation for $Q^{\pi^\text{old}}(s,a)$ to prove that the value function of $\pi^\text{new}$ is better or equal to the value function of $\pi^\text{old}$. 

Let $\pi_\theta(a|s)$ be a parameterization of the policy function that depends on a set of learnable parameters $\theta$. We define the following loss function
\begin{equation}
    L_\pi(\theta) = \mathop{\mathrm{E}}\limits_{{s\sim \mathcal{B} \atop   a\sim \pi_\theta(\cdot|s)}}\left[- Q^{\pi^\text{old}}(s,a) - \alpha_\text{D} H_\text{D}^{\pi_\theta}(s) - \alpha_\text{C} H^{\pi_\theta}_\text{C}(s) \right].
    \label{eq:l_1}
\end{equation}
Thanks to Eqs.~(\ref{eq:z_def}) and (\ref{eq:z_prop}), this choice guarantees us to find a better policy by minimizing $L_\pi(\theta)$ with respect to $\theta$.
In order to evaluate the expectation value in Eq.~(\ref{eq:l_1}), as before we explicitly average over the discrete action and perform a single sample of the continuous action, and we replace $Q^{\pi^\text{old}}$ with $\min_j Q_{\phi_j}$.  Recalling the parameterization in Eq.~(\ref{eq:u_tilda}), this yields
\begin{equation}
\begin{aligned}
&
\begin{multlined}
    L_\pi(\theta) = 
    \underset{ \substack{s\sim \mathcal{B}  }}{\text{E}}\Big[\sum_d \pi_{\text{D},\theta}(d|s) \Big(  \alpha_\text{D} \log\pi_{\text{D},\theta}(d|s) +\\
     \alpha_\text{C} \log\pi_{{\text{C}},\theta}(\tilde{u}_{\theta}(\xi|d,s)|d,s) - 
    \min_{j=1,2} Q_{\phi_j}(s,\tilde{u}_{\theta}(\xi|d,s),d)  \Big) \Big],
    \end{multlined}
    \\ &\quad\quad \xi \sim \mathcal{N}(0,1).
\end{aligned}
    \label{eq:pi_loss}
\end{equation}

We have defined and shown how to evaluate the loss functions $L_Q(\phi)$ and $L_\pi(\theta)$ that allow us to determine the value function and the policy [see Eqs.~(\ref{eq:q_loss}), (\ref{eq:y_2}) and (\ref{eq:pi_loss})]. Now, we discuss how to automatically tune the temperature hyperparameters $\alpha_\text{D}$ and $\alpha_\text{C}$. Ref.~\cite{haarnoja2018_arxiv_sac} shows that constraining the average entropy of the policy to a certain value leads to the same exact SAC algorithm with the addition of an update rule to determine the temperatures. Let $\bar{H}_\text{D}$ and $\bar{H}_\text{C}$ be respectively the fixed average values of the entropy of the discrete and continuous part of the policy. We can then determine the corresponding temperatures $\alpha_\text{D}$ and $\alpha_\text{C}$ minimizing the following two loss functions
\begin{equation}
\begin{aligned}
    L_\text{D}(\alpha_\text{D}) &= \alpha_\text{D}\underset{ \substack{s\sim \mathcal{B} }}{\text{E}}\left[ H^\pi_\text{D}(s) - \bar{H}_\text{D} \right], \\
    L_\text{C}(\alpha_\text{C}) &= \alpha_\text{C}\underset{ \substack{s\sim \mathcal{B} }}{\text{E}}\left[ H^\pi_\text{C}(s) - \bar{H}_\text{C} \right].
\end{aligned}
\end{equation}
As usual, we evaluate the entropies by explicitly taking the average over the discrete actions, and taking a single sample of the continuous action. To be more specific, we evaluate $L_\text{D}$ by computing
\begin{equation}
    L_\text{D}(\alpha_\text{D}) = \alpha_\text{D}\underset{ \substack{s\sim \mathcal{B} }}{\text{E}}\left[ -\sum_d \pi_{\text{D}}(d|s) \log\pi_{\text{D}}(d|s) - \bar{H}_\text{D} \right],
    \label{eq:d_loss}
\end{equation}
and $L_\text{C}$ by computing
\begin{multline}
    L_\text{C}(\alpha_\text{C}) = \alpha_\text{C} \\
    \cdot\underset{ \substack{s\sim \mathcal{B} }}{\text{E}}\left[ -\sum_d \pi_{\text{D}}(d|s) \underset{ \substack{u\sim \pi_{\text{C}}(\cdot|d,s) }}{\text{E}}[ \log\pi_{\text{C}}(u|d,s) ] - \bar{H}_\text{C} \right]
    \label{eq:c_loss}
\end{multline}
and replacing the expectation value over $u$ with a single sample.

To summarize, the SAC algorithm consists of repeating over and over a policy evaluation step, a  policy improvement step, and a step where the temperatures are updated. The policy evaluation step consists of a single optimization step to minimize the loss functions $L_Q(\phi_i)$ (for $i=1,2$), given in Eq.~(\ref{eq:q_loss}), where $y(r,s^\prime)$ is computed using Eq.~(\ref{eq:y_2}).
The policy improvement step consists of a single optimization step to minimize the loss function $L_\pi(\theta)$ given in Eq.~(\ref{eq:pi_loss}). The temperatures are then updated performing a single optimization step to minimize $L_\text{D}(\alpha_\text{D})$ and $L_\text{C}(\alpha_\text{C})$ given respectively in Eqs.~(\ref{eq:d_loss}) and (\ref{eq:c_loss}). In all loss functions, the expectation value over the states is approximated with a batch of experience sampled randomly from the replay buffer $\mathcal{B}$.

We now detail how we parameterize $\pi(a|s)$ and $Q(s,a)$. The idea is to develop an efficient way to process the state that can potentially be a long time-series of actions. To this aim, we introduce a ``convolution block'' as a building element for our NN architecture.
 \begin{figure}[!tb]
	\centering
	\includegraphics[width=0.99\columnwidth]{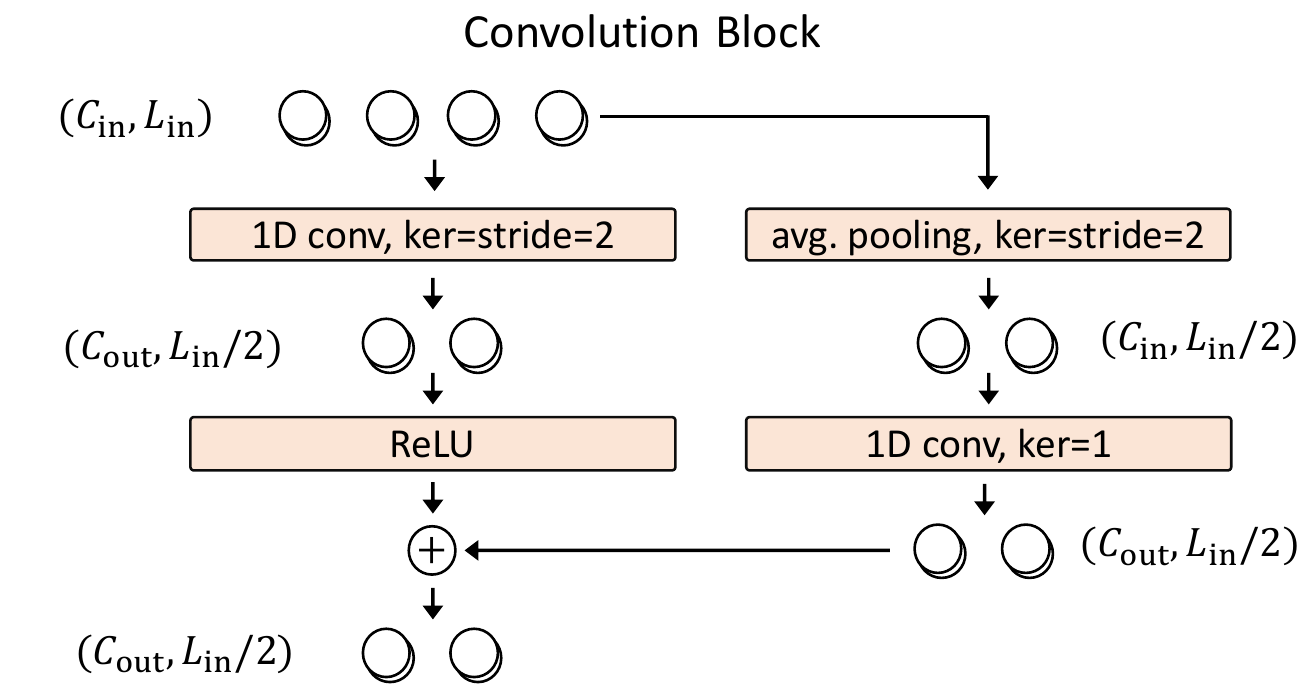}
    \caption{Schematic representation of the convolution block that takes as input a 1D time-series of size $(C_\text{in},L_\text{in})$, where $L_\text{in}$ is the length of the series and $C_\text{in}$ is the number of channels, and produces an output of size $(C_\text{out},L_\text{in}/2)$. In this image $L_\text{in}=4$. The output is produced by stacking a 1D convolution of kernel size and stride of $2$, and a non-linearity (left branch). A residual connection (right branch), consisting only of linear operations, is added to improve trainability. }
	\label{fig:conv_block}
\end{figure}
The convolution block, detailed in Fig.~\ref{fig:conv_block}, takes an input of size $(C_\text{in}, L_\text{in})$, where $C_\text{in}$ is the number of channels (i.e. the number of parameters determining an action at every time-step) and $L_\text{in}$ is the length of the time-series, and produces an output of size $(C_\text{out},L_\text{out}=L_\text{in}/2)$, thus halving the length of the time-series. Notice that we include a skip connection (right branch in Fig.~\ref{fig:conv_block}) to improve trainability \cite{he2015}.

Using the decomposition in Eq.~(\ref{eq:pi_decomp}) and the parameterization in Eq.~(\ref{eq:u_tilda}), the quantities that need to be parameterized are the discrete probabilities $\pi_{\text{D}}(d|s)$, the averages $\mu(d,s)$ and the variances $\sigma(d,s)$, for $d=1,\dots, |D|$, $|D|=3$ being the number of discrete actions. 
 \begin{figure}[!tb]
	\centering
	\includegraphics[width=0.99\columnwidth]{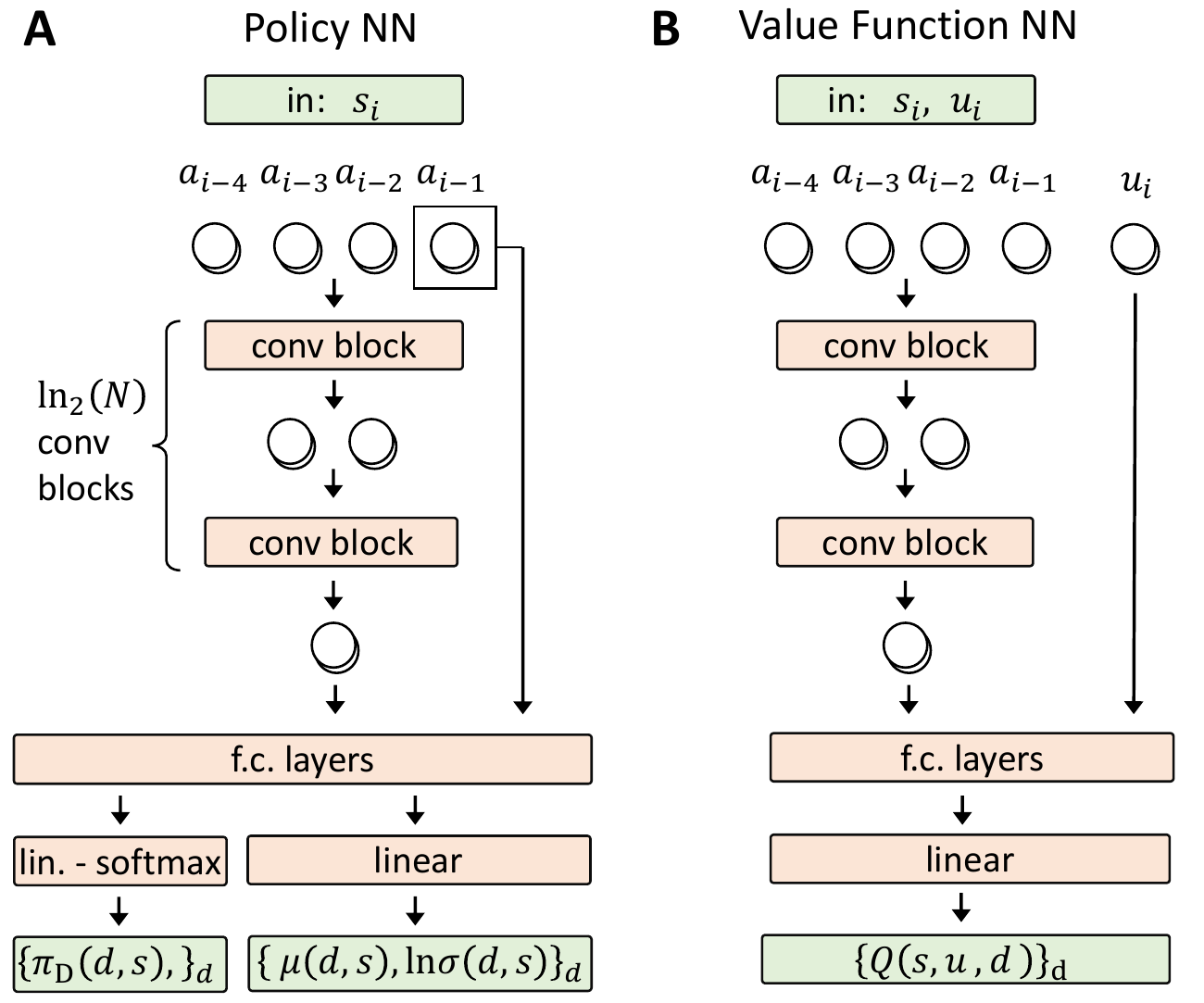}
    \caption{Neural network architecture used to parameterize the policy $\pi(\vec{u},d|s)$ (panel A) and to parameterize the value function $Q(s,\vec{u},d)$ (panel B). }
	\label{fig:nns}
\end{figure}
The architecture of the neural network that we use for the policy function is shown in Fig.~\ref{fig:nns}A. The state, composed of the time-series $s_i=(a_{i-N},\dots, a_{i-1})$ which has shape $(C_\text{in}, L_\text{in}=N)$, is fed through a series of $\ln_2(N)$ convolutional blocks, which produce an output of length $(C_\text{out},L=1)$. The number of input channels $C_\text{in}$ is determined by stacking the components of $\vec{u}$ (which, for simplicity, is a single real number $u$ in this appendix) and by using a one-hot encoding of the discrete actions. We then feed this output, together with the last action which has a privileged position, to a series of fully connected NNs with ReLU activations. Finally, a linear network outputs $W(d|s)$, $\mu(d,s)$ and $\log(\sigma(d,s))$, for all $d=1,\dots, |D|$. The probabilities $\pi_{\text{D}}(d|s)$ are then produced applying the softmax operation to $W(d|s)$. 

We parameterize the value function $Q_\phi(s,u,d)$ as in Fig.~\ref{fig:nns}B. As for the policy function, the state $s$ is fed through $\ln_2(N)$ stacked convolution blocks which reduce the length of the input to $(C_\text{out},L=1)$. This output, together with the action $u$, is fed into a series of fully-connected layers with ReLU activations. We then add a linear layer that produces $|D|$ outputs, corresponding to the value of $Q(s,u,d)$ for each $d=1,\dots,|D|$.

At last, we discuss a further change to the current method that we implemented in the superconducting qubit refrigerator case to improve the converge. This idea is the following. The return $\ev*{r_c}$ is a convex combination of the power and of the negative entropy production. The first term is positive when the system is delivering the desired power, while the second term is strictly negative. Therefore, for $c$ close to $1$, the optimal value of the return is some positive quantity. Instead, as $c$ decreases, the optimal value of the return decreases, getting closer to zero (this can be seen explicitly in Figs.~\ref{fig:qubit_fridge_pareto}A and \ref{fig:harmonic_engine}B). However, a null return can also be achieved by a trivial cycle that consists of doing nothing, i.e. of keeping the control constant in time. Indeed, this yields both zero power, and zero entropy production. Therefore, as $c$ decreases, it becomes harder and harder for the RL agent to distinguish good cycles from these trivial solutions. We thus modify our method to allow us to smoothly change the value of $c$ during training from $1$ to the desired final value, which allows to tackle an optimization problem by ``starting from an easier problem'' ($c=1$), and gradually increasing its difficulty. This required the following modifications to the previously described method.

We introduce two separate value functions, one for each objective (P for the power, and $\Sigma$ for the entropy production)
\begin{equation}
\begin{aligned}
\begin{multlined}
    Q^\pi_\text{P}(s,a) = 
    \text{E}_\pi  \Big[ r_{1}^\text{(P)} + 
    \sum_{k=1}^{\infty} \gamma^k \,\Big(r^{(\text{P})}_{k+1} + \alpha_\text{D} H^\pi_\text{D}(s_k) \\
    + \alpha_\text{C}H^\pi_\text{C}(s_k)  \Big) \Big| s_0=s, a_0=a \Big], 
\end{multlined}
\\
\begin{multlined}
    Q^\pi_\Sigma(s,a) = 
    \text{E}_\pi  \Big[ r_{1}^{(\Sigma)} + 
    \sum_{k=1}^{\infty} \gamma^k \,\Big(r^{(\Sigma)}_{k+1} + \alpha_\text{D} H^\pi_\text{D}(s_k) \\
    + \alpha_\text{C}H^\pi_\text{C}(s_k)  \Big) \Big| s_0=s, a_0=a \Big],
\end{multlined}
\end{aligned}
\label{eq:two_values}
\end{equation}
where 
\begin{equation}
\begin{aligned}
	r_{i+1}^\text{(P)} &\equiv \frac{1}{\Delta t} \int\limits_{t_i}^{t_i+\Delta t} \frac{P(\tau)}{P_0}  \, d\tau, &
	r_{i+1}^{(\Sigma)} &\equiv \frac{1}{\Delta t} \int\limits_{t_i}^{t_i+\Delta t} \frac{\Sigma(\tau)}{\Sigma_0} \, d\tau,
\end{aligned}
\label{eq:two_rewards}
\end{equation}
represent respectively the normalized average power and average entropy production during each time-step. Since the value functions in Eq.~(\ref{eq:two_values}) are identical to Eq.~(\ref{eq:q_def}) up to a change of the reward, they separately satisfy the same Bellmann equation as in Eq.~(\ref{eq:bellman}), with $r_1$ replaced respectively with $r_1^\text{(P)}$ and $r_1^{(\Sigma)}$. Therefore, we learn each value functions minimizing the same loss function $L_Q$ given in Eq.~(\ref{eq:q_loss}), with $r_i$ replaced with $r_1^\text{(P)}$ or $r_1^{(\Sigma)}$. Both value functions are parameterized using the same architecture, but separate and independent parameters. We now turn to the determination of the policy. Comparing the definition of $r_{i}$ given in the main text with Eq.~(\ref{eq:two_rewards}), we see that $r_{i+1}=c r_{i+1}^\text{(P)} - (1-c)r_{i+1}^{(\Sigma)}$. Using this property, and comparing Eq.~(\ref{eq:q_def}) with Eq.~(\ref{eq:two_values}), we see that
\begin{equation}
    Q^\pi(s,a) = c Q^\pi_\text{P}(s,a) - (1-c)Q^\pi_{\Sigma}(s,a).
    \label{eq:q_two_objevtives}
\end{equation}
Therefore, we learn the policy minimizing the same loss function as in Eq.~(\ref{eq:pi_loss}), using Eq.~(\ref{eq:q_two_objevtives}) to compute the value function.
To summarize, this method allows us to vary $c$ dynamically during training. This requires learning two value functions, one for each objective, and storing in the replay buffer the two separate rewards $r_i^\text{(P)}$ and $r_i^{(\Sigma)}$.

At last, when we refer to ``final deterministic cycle'', we are sampling from the policy function ``switching off the stochasticity'', i.e. choosing continuous actions $u$ setting $\xi=0$ in Eq.~(\ref{eq:u_tilda}), and choosing deterministically the discrete action with the highest probability.

~\\ \textbf{Physical model}

As discussed in the main text, we describe the dynamics of the two analyzed QTMs employing the Lindblad master equation that can be derived also for non-adiabatic drivings \cite{yamaguchi2017}, in the weak system-bath coupling regime performing the usual Born-Markov and secular approximation \cite{gorini1976,lindblad1976,breuer2002} and neglecting the Lamb-shift contribution.
This approach describes the time-evolution of the reduced density matrix of the quantum system, $\hat{\rho}(t)$, under the assumption of weak system-bath interaction. Setting $\hbar=1$, the master equation reads
\begin{equation}
	\frac{\partial }{\partial t} {\hat{\rho}}(t) = -i\left[ \hat{H}[\vec{u}(t)], \hat{\rho}(t)\right] + \sum\nolimits_{\alpha} \mathcal{D}^{(\alpha)}_{\vec{u}(t),d(t)}[\hat{\rho}(t)],
	\label{eq:lindblad}
\end{equation}
where $\hat{H}[\vec{u}(t)]$ is the Hamiltonian of the quantum system that depends explicitly on time via the control parameters $\vec{u}(t)$, $[\cdot,\cdot]$ denotes the commutator, and $\mathcal{D}^{(\alpha)}_{\vec{u}(t),d(t)}[\cdot]$, known as the dissipator, describes the effect of the coupling between the quantum system and bath $\alpha = \mathrm{H}, \mathrm{C}$. 
 We notice that since the RL agent produces piece-wise constant protocols, we are not impacted by possible inaccuracies of the master equation subject to fast parameter driving \cite{dann2018}, provided that $\Delta t$ is not smaller than the bath timescale.  
Without loss of generality, the dissipators can be expressed as \cite{lindblad1976,breuer2002}
\begin{multline}
	\mathcal{D}^{(\alpha)}_{\vec{u}(t),d(t)} = \lambda_\alpha[d(t)] \sum_k \gamma^{(\alpha)}_{k,\vec{u}(t)}
	\Big( \hat{A}^{(\alpha)}_{k,\vec{u}(t)} \hat{\rho}\hat{A}^{(\alpha)\dagger}_{k,\vec{u}(t)} \\ 
	-\frac{1}{2}\hat{A}^{(\alpha)\dagger}_{k,\vec{u}(t)} \hat{A}^{(\alpha)}_{k,\vec{u}(t)} \hat{\rho} -\frac{1}{2} \hat{\rho}\hat{A}^{(\alpha)\dagger}_{k,\vec{u}(t)} \hat{A}^{(\alpha)}_{k,\vec{u}(t)}\Big),
\end{multline}
where $\lambda_\alpha[d(t)] \in \{0,1\}$ are functions that determine which bath is coupled the quantum system, $\hat{A}^{(\alpha)}_{k,\vec{u}(t)}$ are the Lindblad operators, and $\gamma^{(\alpha)}_{k,\vec{u}(t)}$ are the corresponding rates.
In particular, $\lambda_\text{H}(\text{Hot}) = 1$, $\lambda_\text{C}(\text{Hot})=0$, while $\lambda_\text{H}(\text{Cold}) = 0$, $\lambda_\text{C}(\text{Cold})=1$, and $\lambda_\text{H}(\text{None})=\lambda_\text{C}(\text{None}) = 0$.
Notice that both the Lindblad operators and the rates can depend on time through the value of the control $\vec{u}(t)$. Their explicit form depends on the details of the system, i.e. on the Hamiltonian describing the dynamics of the \textit{overall system} including the bath and the system-bath interaction. Below, we provide the explicit form of $\hat{A}^{(\alpha)}_{k,\vec{u}(t)}$ and $\gamma^{(\alpha)}_{k,\vec{u}(t)}$ used to model the two setups considered in the manuscript. 
We adopt the standard approach to compute the instantaneous power and heat currents \cite{alicki1979}
\begin{equation}
\begin{aligned}
	P(t) &\equiv -\Tr\left[\hat{\rho}(t)\, \frac{\partial}{\partial t}\hat{H}[\vec{u}(t)]\right] , \\
	J_\alpha(t) &\equiv \Tr\left[{\hat{H}}[\vec{u}(t)]\, \mathcal{D}^{(\alpha)}_{\vec{u}(t),d(t)}\right],
\end{aligned}
\end{equation}
that guarantees the validity of the first law of thermodynamics $\partial{U}(t)/(\partial t) = -P(t) + \sum_\alpha J_\alpha(t)$, the internal energy being defined as $U = \mathrm{Tr}[\hat{\rho}(t)\hat{H}[\vec{u}(t)]]$.

In the superconducting qubit refrigerator, we employ the model first put forward in Ref.~\cite{karimi2016}, and further studied in Refs.~\cite{pekola2019,funo2019}. In particular, we consider the following Lindblad operators and corresponding rates (identifying $k= \pm$):
\begin{equation}
\begin{aligned}
	\hat{A}^{(\alpha)}_{+,u(t)} &= -i\ket*{e_{u(t)}}\bra*{g_{u(t)}}, &
	\hat{A}^{(\alpha)}_{-,u(t)} &= +i\ket*{g_{u(t)}}\bra*{e_{u(t)}},
\end{aligned}
\end{equation}
where $\ket*{g_{u(t)}}$ and $\ket*{e_{u(t)}}$ are, respectively, the instantaneous ground state and excited state of Eq.~(\ref{eq:h_fridge}). The corresponding rates are given by $\gamma^{(\alpha)}_{\pm,u(t)} = S_{\alpha}[\pm\Delta \epsilon_{u(t)}] $, where $\Delta \epsilon_{u(t)}$ is the instantaneous energy gap of the system, and
\begin{equation}
	S_\alpha(\Delta \epsilon)= \frac{g_{\alpha}}{2} \frac{1}{1+Q_\alpha^2( \Delta\epsilon/\omega_\alpha - \omega_\alpha/\Delta \epsilon )^2 } \frac{\Delta \epsilon}{e^{\beta_\alpha\Delta\epsilon}-1}
\end{equation}
is the noise power spectrum of bath $\alpha$. Here $\omega_\alpha$, $Q_\alpha$ and $g_\alpha$ are the base resonance frequency, quality factor and coupling strength of the resonant circuit acting as bath $\alpha=\text{H},\text{C}$ (see Refs.~\cite{karimi2016,funo2019} for details). As in Ref.~\cite{funo2019}, we choose $\omega_\text{C}=2E_0\Delta$ and $\omega_\text{H}=2E_0\sqrt{\Delta^2 +1/4}$, such that the C (H) bath is in resonance with the qubit when $u=0$ ($u=1/2$). The width of the resonance is governed by $Q_\alpha$. 
The total coupling strength to bath $\alpha$, plotted in Fig.~\ref{fig:qubit_fridge}F, is quantified by 
\begin{equation}
	\gamma^{(\alpha)}_{u(t)} \equiv \gamma^{(\alpha)}_{+,u(t)} + \gamma^{(\alpha)}_{-,u(t)}.
\end{equation}

In the quantum harmonic oscillator based heat engine, following Ref.~\cite{rezek2006}, we describe the coupling to the baths through the Lindblad operators  $\hat{A}^{(\alpha)}_{+,u(t)} = \hat{a}_{u(t)}^\dagger$, $\hat{A}^{(\alpha)}_{-,u(t)} = \hat{a}_{u(t)}$ and corresponding rates $\gamma^{(\alpha)}_{+,u(t)} = \Gamma_\alpha \,n(\beta_\alpha u(t)\omega_0)$ and $\gamma^{(\alpha)}_{-,u(t)} = \Gamma_\alpha[1+ n(\beta_\alpha u(t) \omega_0 )]$, where we identify $k= \pm$. $\hat{a}_{u(t)}=(1/\sqrt{2})\sqrt{m\omega_0 u(t)}\,\hat{q} + i/\sqrt{m\omega_0 u(t)}\,\hat{p}$ and $\hat{a}_{u(t)}^\dagger$ are respectively the (control dependent) lowering and raising operators, $\Gamma_{\alpha}$ is a constant rate setting the thermalization timescale of the system coupled to bath $\alpha$, and $n(x)=[\exp(x)-1]^{-1}$ is the Bose-Einstein distribution.

~\\ \textbf{Training details}

We now provide additional practical details and the hyper parameters  used to produce the results of this manuscript. 

In order to enforce sufficient exploration in the early stage of training, we do the following. As in Ref. \cite{spinningup2018}, for a fixed number of initial steps, we choose random actions sampling them uniformly withing their range. Furthermore, for another fixed number of initial steps, we do not update the parameters to allow the replay buffer to have enough transitions. $\mathcal{B}$ is a first-in-first-out buffer, of fixed dimension, from which batches of transitions $(s_k,a_k,r_{k+1},s_{k+1},a_{k+1})$ are randomly sampled to update the NN parameters. After this initial phase, we repeat a policy evaluation, a policy improvement step  and a temperature update step $n_\text{updates}$ times every $n_\text{updates}$ steps. This way, the overall number of updates coincides with the number of actions performed on the QTM. The optimization steps for the value function and the policy are performed using the ADAM optimizer with the standard values of $\beta_1$ and $\beta_2$. The temperature parameters $\alpha_\text{D}$ and $\alpha_\text{C}$ instead are determined using stochastic gradient descent with learning rate $0.001$. To favor an exploratory behavior early in the training, and at the same time to end up with a policy that is approximately deterministic, we schedule the target entropies $\bar{H}_\text{C}$ and $\bar{H}_\text{D}$. In particular, we vary them exponentially during each step according to
\begin{multline}
    \bar{H}_a(n_\text{steps}) = \bar{H}_{a,\text{end}}  \\
    + (\bar{H}_{a,\text{start}}-\bar{H}_{a,\text{end}})\exp(-n_\text{steps}/\bar{H}_{a,\text{decay}}),
\end{multline}
where $a=\text{C},\text{D}$, $n_\text{steps}$ is the current step number, and $\bar{H}_{a,\text{start}}$, $\bar{H}_{a,\text{end}}$ and $\bar{H}_{a,\text{decay}}$ are hyperparameters. In the superconducting qubit refrigerator case, we schedule the parameter $c$ according to a Fermi distribution, i.e.
\begin{equation}
    c(n_\text{step}) = c_\text{end} + (c_\text{start}-c_\text{end})\left[ 1 + \exp\left( \frac{n_\text{step}-c_\text{mean}}{c_\text{decay}} \right) \right]^{-1}.
    \label{eq:c_fermi}
\end{equation}

In the harmonic oscillator engine case, to improve stability while training for lower values of $c$, we do not vary $c$ during training, as we do in the superconducting qubit refrigerator case. Instead, we discourage the agent from never utilizing one of the two thermal baths by adding a negative reward if, withing the last $N=128$ actions describing the state, less than $25$ describe a coupling to either bath. In particular, if the number of actions $N_\alpha$  where $d=\alpha$, with $\alpha=\text{Hot},\text{Cold}$ is less than 25 in the state time-series, we sum to the reward the following penalty
\begin{equation}
	r_{penalty} = -1.4\frac{25-N_\alpha}{25}.
\end{equation}
This penalty has no impact on the final cycles where $N_\alpha$ is much larger than $25$. 

All hyperparameters used to produce the results of the superconducting qubit refrigerator and of the harmonic oscillator heat engine are provided respectively in Tables~\ref{tab:qubit_hyper} and \ref{tab:harmonic_hyper}, where $c$ refers to the weight at which we are optimizing the return. 
\begin{table}[h]
\centering
\begin{tabular}{lccc}
\toprule
Hyperparameter ~ & ~ Qubit Refrigerator  \\
\midrule
Batch size  & 512  \\
Training steps & 500k \\
learning rate & 0.0003   \\
$\mathcal{B}$ size  & 280k   \\
$\rho_\text{polyak}$ & 0.995  \\
Channels per conv. block ~  & (64, 64, 64, 128, 128, 128, 128)  \\
Units per f.c. layer in $\pi$ & (256) \\
Units per f.c. layer in $Q^\pi$ & (256, 256)  \\
Initial random steps  & 5k \\
First update at step  & 1k  \\
$n_\text{updates}$  & 50   \\
$\bar{H}_{\text{C},\text{start}}$ & 0  \\
$\bar{H}_{\text{C},\text{end}}$ & -3.5  \\
$\bar{H}_{\text{C},\text{decay}}$ & 440k  \\
$c_\text{start}$ & 1 \\
$c_\text{end}$ & $c$  \\
$c_\text{mean}$ & 170k  \\
$c_\text{decay}$ & 20k  \\
\bottomrule
\end{tabular}
\caption{Hyperparameters used in numerical calculations relative to the superconducting qubit refrigerator that are not reported in the caption of Fig.~\ref{fig:qubit_fridge}.}
\label{tab:qubit_hyper}
\end{table}
\begin{table}[h]
\centering
\begin{tabular}{lccc}
\toprule
Hyperparameter ~ &   ~ Harmonic Engine  \\
\midrule
Batch size  & 512 \\
Training steps & 500k \\
learning rate  & 0.0003  \\
$\mathcal{B}$ size  & 160k  \\
$\rho_\text{polyak}$  & 0.995 \\
Channels per conv. block  ~ & ~ (64, 64, 64, 128, 128, 128, 128) \\
Units per f.c. layer in $\pi$ & (256) \\
Units per f.c. layer in $Q^\pi$ & (256, 128) \\
Initial random steps  &  5k \\
First update at step  & 1k \\
$n_\text{updates}$  & 50  \\
$\bar{H}_{\text{C},\text{start}}$  & -0.72 \\
$\bar{H}_{\text{C},\text{end}}$  & -3.5 \\
$\bar{H}_{\text{C},\text{decay}}$  & 144k \\
$\bar{H}_{\text{D},\text{start}}$ & $\ln 3$ \\
$\bar{H}_{\text{D},\text{end}}$  & 0.01 \\
$\bar{H}_{\text{D},\text{decay}}$ & 144k \\
\bottomrule
\end{tabular}
\caption{Hyperparameters used in numerical calculations relative to the harmonic oscillator heat engine that are not reported in the caption of Fig.~\ref{fig:harmonic_engine}.}
\label{tab:harmonic_hyper}
\end{table}

~\\ \textbf{Convergence of the RL approach}
The training process presents some degree of stochasticity, such as the initial random steps, the stochastic sampling of actions from the policy function, and the random sampling of a batch of experience from the replay buffer to compute an approximate gradient of the loss functions. We thus need to evaluate the reliability of our approach.  

As shown in the main text, specifically in Figs.~\ref{fig:qubit_fridge_pareto} and \ref{fig:harmonic_engine}, we ran the full optimization $5$ times. Out of $65$ trainings in the superconducting qubit refrigerator case, only $4$ failed, and out of the $55$ in the harmonic oscillator engine, only $2$ failed, where by failed we mean that the final return was negative. In such cases, we ran the training an additional time. 

Figs.~\ref{fig:qubit_fridge_pareto}A and \ref{fig:harmonic_engine}B display an error bar corresponding to the standard deviation, at each value of $c$, computed over the $5$ repetitions. Instead, in Figs.~\ref{fig:qubit_fridge_pareto}B and \ref{fig:harmonic_engine}C we display one black dot for each individual training. As we can see, the overall performance is quite stable and reliable.

\begin{figure}[!tb]
	\centering
	\includegraphics[width=0.99\columnwidth]{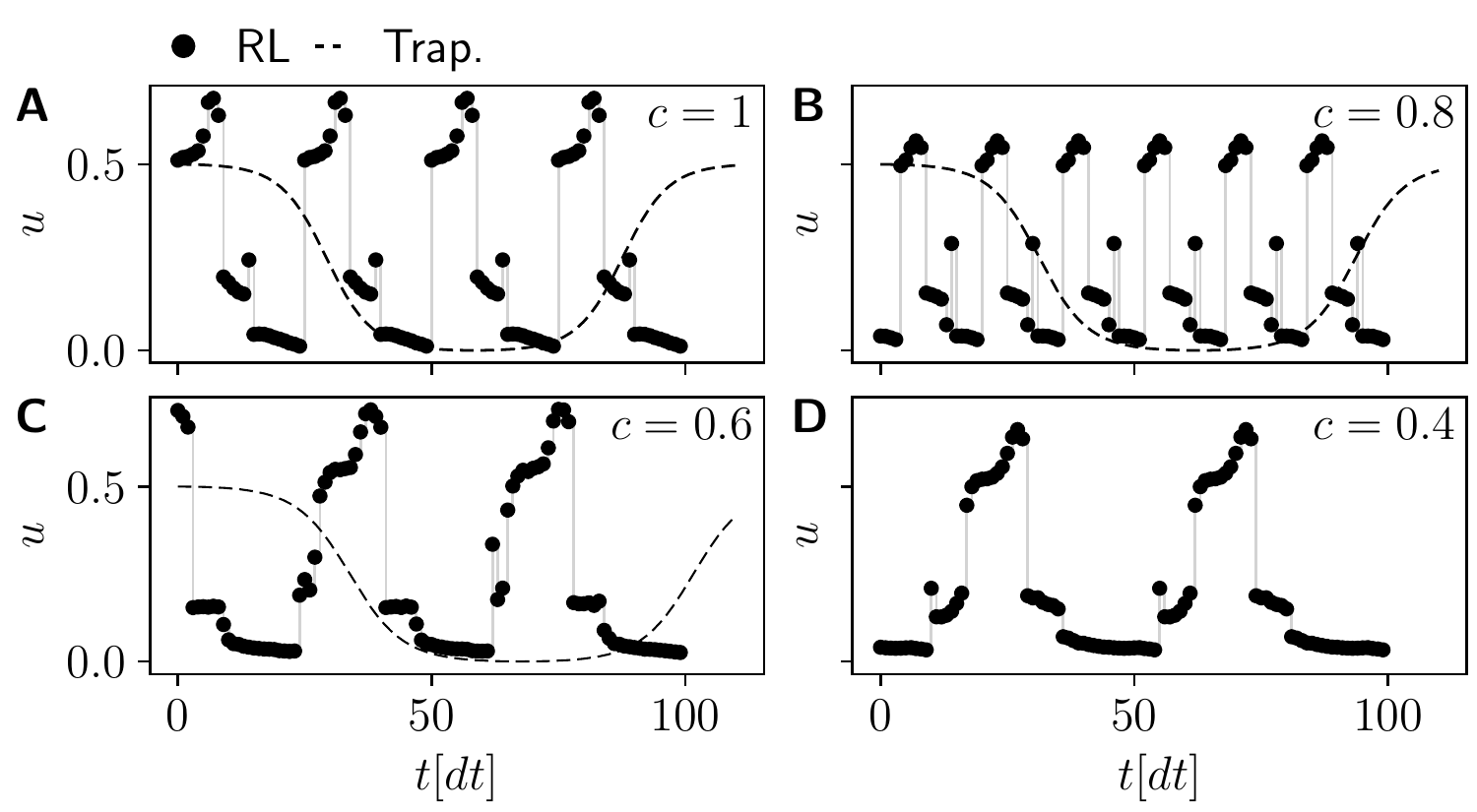}
	\caption{Final deterministic cycle, identified in the superconducting qubit refrigerator, at the fifth training. Same parameters and quantities are shown as in Figs.~\ref{fig:qubit_fridge_pareto}C-F.}
	\label{fig:qubit_extra_cycles}
\end{figure}
\begin{figure}[!tb]
	\centering
	\includegraphics[width=0.99\columnwidth]{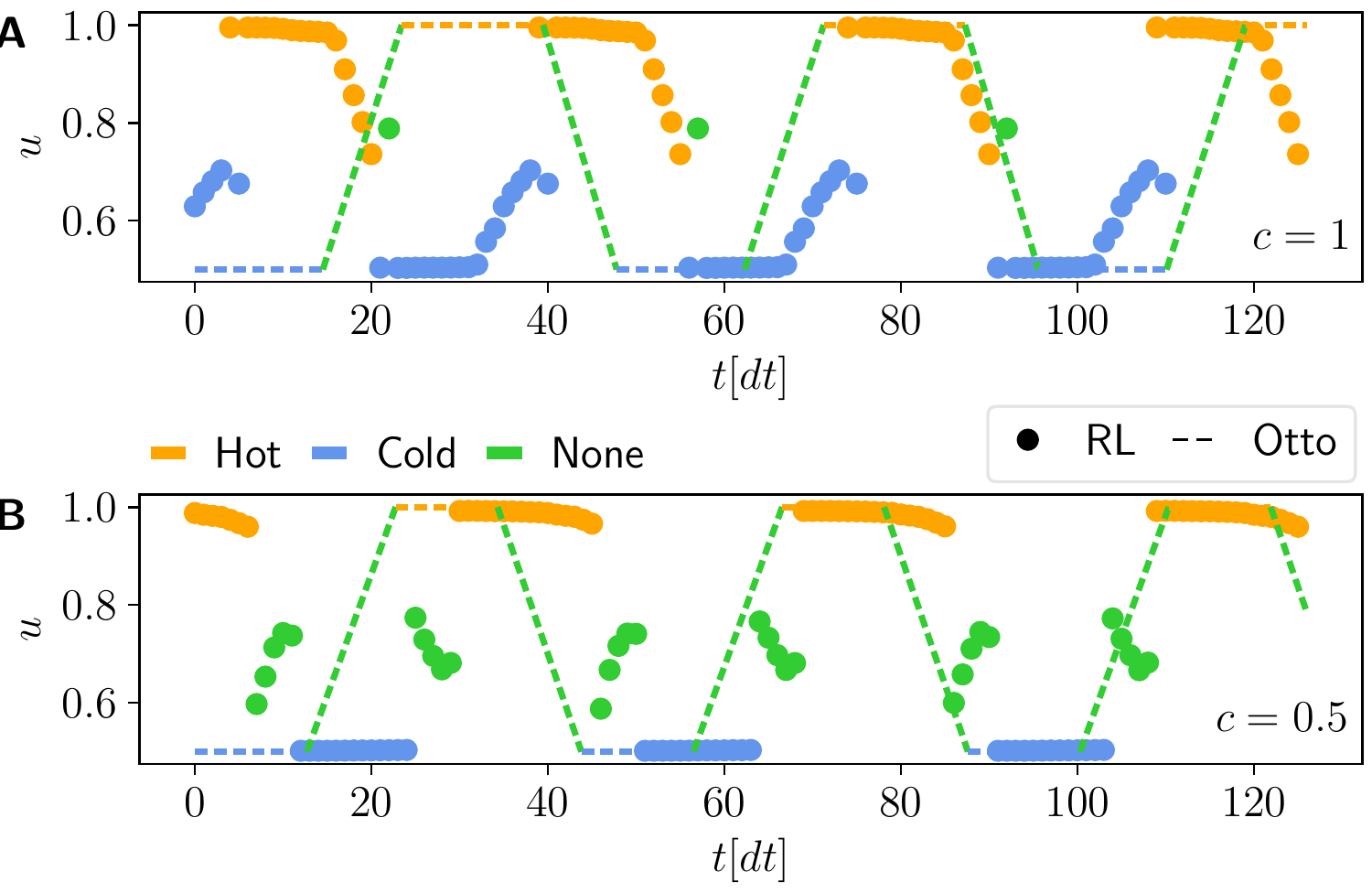}
	\caption{Final deterministic cycle, identified in the harmonic oscillator engine, at the fifth training. Same parameters and quantities are shown as in Figs.~\ref{fig:harmonic_engine}D-E. }
	\label{fig:harmonic_extra_cycles}
\end{figure}
At last, we discuss the variability of the discovered cycles. The cycles shown in Figs. \ref{fig:qubit_fridge_pareto}C-F and \ref{fig:harmonic_engine}D-E were chosen by selecting the largest return among the $5$ repetitions. In Figs.~\ref{fig:qubit_extra_cycles} and \ref{fig:harmonic_extra_cycles} we display cycles discovered in the last of the $5$ repetition, i.e. chosen without any post-selection. They correspond to the same setups and parameters displayed in Figs. \ref{fig:qubit_fridge_pareto}C-F and \ref{fig:harmonic_engine}D-E. As we can see, $5$ out of the $6$ displayed cycles are very similar to the ones displayed in Figs. \ref{fig:qubit_fridge_pareto}C-F and \ref{fig:harmonic_engine}D-E, with a very slight variability. The only exception is Fig.~\ref{fig:qubit_extra_cycles}B, where the cycle has a visibly shorter period and amplitude than the one shown in Fig.~\ref{fig:qubit_fridge_pareto}D. Despite this visible difference in the cycle shape, the return of the cycle shown in Fig.~\ref{fig:qubit_extra_cycles}B is $0.382$ compared to $0.385$ of the cycle shown in Fig.~\ref{fig:qubit_fridge_pareto}B.

We therefore conclude that, up to minor changes, the cycles are generally quite stable across multiple trainings.

~\\ \textbf{Comparing with other methods}

In Figs.~\ref{fig:qubit_fridge_pareto} and \ref{fig:harmonic_engine} we compare the performance of our method respectively against optimized trapezoidal cycles, and optimized Otto cycles. In both cases, we also maximize the power using the RL method of Ref.~\cite{erdman2022}. We now detail how we perform such comparison.

In the refrigerator based on a  superconducting qubit, we consider the trapezoidal cycle proposed in Ref.~\cite{karimi2016,funo2019}, i.e. we fix
\begin{equation}
    u(t) = \frac{1}{4}\left( 1 + \frac{\tanh(a\,\cos\Omega t)}{\tanh(a)}  \right)
\end{equation}
with $a=2$, and we optimize $\ev{r_c}$ with respect to frequency $\Omega$. In the heat engine case based on a quantum harmonic oscillator, we fix an Otto cycle as described in Ref.~\cite{rezek2006}, i.e. a trapezoidal cycle consisting of the 4 strokes shown in Figs.~\ref{fig:harmonic_engine}D-E as a dashed line, and we optimize over the duration of each of the 4 strokes. In particular, we first performed a grid search in the space of these four durations for $c=1$. After identifying the largest power, we ran the Newton algorithm to further maximize the return. We then ran the Newton algorithm for all other values of $c$. 

The comparison with Ref.~\cite{erdman2022} was done using the source code provided in Ref.~\cite{erdman2022}, and using the same exact hyperparameters that were used in Ref.~\cite{erdman2022}. 

In particular, in the case of the refrigerator based on a superconducting qubit, we re-ran the code using the hyperparameters reported in Table 1, column ``Figs. 3, 4'', of the Methods section of Ref.~\cite{erdman2022}, and we trained for the same number of steps (500k). We then evaluated its power and coefficient of performance evaluating the deterministic policy (which typically has a better performance).
In the heat engine case based on a quantum harmonic oscillator, we evaluated the performance of the cycle reported in Fig.~5a,c of Ref.~\cite{erdman2022}, whose training hyperparameters are reported in Table 1, column ``Fig. 5a'', of the Methods section of Ref.~\cite{erdman2022}.

~\\ \textbf{Generation of coherence}

In order to quantify the coherence generated in the instantaneous eigenbasis of the Hamiltonian in the refrigerator based on a superconducting qubit, we evaluated the time average of \textit{relative entropy of coherence} \cite{baumgratz2014}, defined as
\begin{equation}
	C(\hat{\rho}(t)) = S(\hat{\rho}_\mathrm{diag.}(t)) -  S(\hat{\rho}(t)),
\end{equation}
where $S(\hat{\rho}) = -\mathrm{Tr}[\hat{\rho}\ln\hat{\rho}]$ is the Von Neumann entropy, and
\begin{multline}
	\hat{\rho}_\mathrm{diag.}(t) = \langle g_{u(t)} | \hat{\rho}(t)| g_{u(t)}\rangle \cdot | g_{u(t)}\rangle \langle g_{u(t)} | \\ +  \langle e_{u(t)} | \hat{\rho}(t)| e_{u(t)}\rangle \cdot | e_{u(t)}\rangle \langle e_{u(t)} |
\end{multline}
is the density matrix, in the instantaneous eigenbasis $|g_{u(t)}\rangle$ and $|e_{u(t)}\rangle$, with the off-diagonal terms canceled out. 

We compute the time-average of the relative entropy of coherence generated by the final deterministic cycle found by the RL agent, and compare it to the coherence generated by a trapezoidal cycle operated at the same speed, i.e. with the same period. As we can see in Table~\ref{tab:coherence}, the trapezoidal cycles generate twice as much coherence as the RL cycles shown in Figs.~\ref{fig:qubit_fridge_pareto}C-F, i.e. corresponding to $c=1, 0.8, 0.6, 0.4$.
\begin{table}[h]
\centering
\begin{tabular}{lccc}
\toprule
$c$ ~~ &  ~~ RL ~~ & ~~ Trapez. \\
\midrule
1 & 0.068 & 0.13 \\
0.8 & 0.050 & 0.12 \\
0.6 & 0.054 & 0.092 \\
0.4 & 0.035 & 0.090 \\
\bottomrule
\end{tabular}
\caption{Coherence generated by the final deterministic cycles identified by the RL method (RL column) and generated by a trapezoidal cycle operated at the same speed (Trapez. column) at the values of $c$ shown in the first column. These values correspond to the cycles shown in Figs.~\ref{fig:qubit_fridge_pareto}C-F.}
\label{tab:coherence}
\end{table}

\section*{Code and data availability}
The code used to generate all results is available on GitHub (\url{https://github.com/PaoloAE/paper_rl_blackbox_thermal_machines}).
All raw data that was generated with the accompanying code and that was used to produce the results in the manuscript is available on Figshare
(\url{https://doi.org/10.6084/m9.figshare.19180907}).

\section*{Acknowledgements} We are greatly thankful to Mart{\'i} Perarnau-Llobet, Paolo Abiuso and Alberto Rolandi for useful discussions and for suggesting to include the entropy production in the return. We gratefully acknowledge funding by the BMBF (Berlin Institute for the Foundations of Learning and Data -- BIFOLD), the European Research Commission (ERC CoG 772230) and the Berlin Mathematics Center MATH+ (AA1-6, AA2-8, AA2-18).

\section*{Competing interests}
The authors declare no competing interests.
P.A.E. and F.N. are authors of a patent application containing aspects of this work
(Application to the European Patent Office, file number: 21 191 966.7).

\section*{Author Contributions}
P.A.E. and F.N. designed the research and method. P.A.E. wrote the computer code and carried out the numerical calculations. P.A.E. and F.N. analysed the data and wrote the manuscript.

\clearpage

\bibliography{references}

\end{document}